\documentclass[12pt,preprint]{emulateapj}

\newcommand{\fw}{\filter{F160W}}
\newcommand{\h}{\filter{H}}
\newcommand{\filter}[1]{\mbox{\it #1\/}}              

\shorttitle{The low-mass IMF in the 30 Doradus cluster}
\shortauthors{Andersen et al.}

\begin{document}

\title{The low-mass IMF in  the 30 Doradus starburst cluster}


\author{M. Andersen}
\affil{Space Science Department, European Space Agency, Keplerlaan 1, 2200 AG Noordwijk, Netherlands}
\email{manderse@rssd.esa.int}

\author{H. Zinnecker} 
\affil{Astrophysical Institute Potsdam, An der Sternwarte 16, 14482 Potsdam Germany}
\author{A. Moneti}
\affil{Institut d'Astrophysique, Paris 98bis Blvd Arago, F-75014 Paris, France }
\author{M. J. McCaughrean}
\affil{Space Science Department, European Space Agency, Keplerlaan 1, 2200 AG }

\author{B. Brandl}
\affil{Leiden Observatory, P.O. Box 9513, 2300 RA Leiden, Netherlands}

\author{W. Brandner}
\affil{Max-Planck-Institut f\"ur Astronomie, K\"onigstuhl 17, 69117 Heidelberg, Germany}

\author{G. Meylan}
\affil{Laboratoire d'Astrophysique, Ecole Polytechnique F\'ed\'erale de Lausanne (EPFL), Observatoire, CH - 1290 Sauverny, Switzerland}
\and
\author{D. Hunter}
\affil{Lowell Observatory, 1400 West Mars Hill Road, Flagstaff, AZ 86001}




\begin{abstract}
 We present  deep Hubble Space Telescope (HST) NICMOS 2 \filter{F160W} band observations  of the central 56\arcsec$\times$57\arcsec\ (14pc$\times$14.25pc) region around R136  in  the  starburst  cluster 30 Dor (NGC 2070) located in the Large Magellanic Cloud. 
Our aim is to derive the stellar Initial Mass Function (IMF)  down to $\sim$1 M$_\odot$ in order to test whether the IMF in a massive metal--poor cluster is similar to that observed in nearby young clusters and the field  in our Galaxy. 
We estimate the mean age of the cluster  to be 3 Myr by combining our \filter{F160W} photometry with previously obtained HST WFPC2 optical \filter{F555W} and \filter{F814W}  band photometry  and  comparing the stellar locus in the color--magnitude diagram with  main sequence and pre--main sequence isochrones. 
The color--magnitude diagrams show the presence of differential extinction and  possibly  an  age spread of 
a few Myr. 
 We convert the magnitudes into masses adopting both a single mean age of  
3 Myr isochrone and  a constant star formation history from 2 to 4 Myr. 
We derive the  IMF after correcting for incompleteness due to crowding. 
The faintest stars detected have a mass of 0.5 M$_\odot$ and the data are more 
than  50\%\ complete outside a radius of  5 pc down to a mass limit of  
1.1 M$_\odot$ for 3 Myr old objects.    
We find an IMF  of $\frac{dN}{d\log M }\propto M^{-1.20\pm 0.2}$ over the mass 
range 1.1--20 M$_\odot$ only slightly shallower than a Salpeter IMF\@. 
In particular, we find no strong evidence for a flattening of the IMF down 
to 1.1 M$_\odot$ at a distance of 5 pc from the center, in contrast to a 
flattening at 2 M$_\odot$ at a radius of 2 pc, reported in a previous optical
 HST study. 
flattening at 2 M$_\odot$ at a radius of  2 pc previously found. 
We examine several possible reasons for the different results including the 
possible presence of mass segregation and the effects of differential 
extinction, particularly for the pre--main sequence sources. 
If the IMF determined here applies to the whole cluster, the cluster would be massive enough to 
remain bound and evolve into a relatively low--mass globular cluster.

\end{abstract}





\section{Introduction}
The shape of the stellar Initial Mass Function (IMF) and whether it is 
universal or not are key issues  in astrophysics. 
For clusters within 2 kpc, there is no  compelling evidence  for variations in 
the stellar IMF \citep[e.g.][]{meyer00,kro02,chabrier_conf} or the brown dwarf 
IMF \citep[e.g.][]{andersen08}. 
However, these clusters only span a limited range in total cluster mass 
($10^2-10^3$ M$_\odot$) and all  have a metallicity similar to the solar value. 
Thus, we are forced to observe more extreme regions of star formation in 
search of  variations in the IMF as a function of environment.
It has been suggested that the shape of the IMF and in particular the 
characteristic mass where the IMF flattens from a Salpeter power--law  
could depend on the  metallicity  in the  molecular cloud out of which the 
stars are formed. 
\citet{low}, \citet{larson}, and \citet{omukai} suggest that a lower 
metallicity results in higher temperatures in the molecular cloud which 
would increase the Jeans mass. 
This would in turn result in a top heavy IMF relative to the solar
 metallicity IMF\@. 


The closest place with massive metal--poor young star clusters is the 
Large Magellanic Cloud (LMC). 
The metallicity is only $\frac{1}{3}-\frac{1}{2}$  the solar value 
\citep{smith}  and   star clusters can be studied in some detail despite a 
distance of $\sim$50 kpc \citep{westerlund}.  
Of particular interest is the 30 Dor cluster which is powering the most 
luminous HII region in the Local Group \citep{kennicutt}. 
The cluster has a mass of at least 2.2$\times 10^4$ M$_\odot$ within a  radius 
of 4.7 pc \citep{hunter95} and is a relatively low-mass analog to the more 
distant starburst clusters. 
R136 lies at the center of the 30 Dor cluster and has long commanded 
significant attention: Once thought to be a single $\sim$1000 M$_\odot$ star 
\citep{cassinelli}, the region is now known to host  numerous O  stars  
\citep{melnick85,weigelt,pehlemann,campbell}. 

 The  whole   30 Dor region, with a size of 200 pc, appears to have an age 
spread of $\sim$20 Myr \citep{mcgregor,selman2} with stars still forming 
\citep{rubio,maercker}.  
R136 appears to have a much smaller age spread of at most a   few Myr 
\citep{melnick85,brandl96,masseyhunter}. 
 An  age of  2 Myr or less  is inferred from   spectroscopy of the O stars in 
the very cluster center \citep{masseyhunter}, whereas the  intermediate mass 
population is thought  to be  $\sim$3--4 Myr old \citep{hunter95}. 

\citet{masseyhunter}  obtained   HST spectroscopy  of the 65 bluest and most 
luminous sources within 17\arcsec\ of the cluster center. 
They derived the IMF over the mass range 15--120 M$_\odot$ and found it  to be 
well approximated by a power--law $\frac{dN}{dlog M}\propto M^{\Gamma}$ with a 
slope of $\Gamma=-1.3\pm0.1$, consistent with a Salpeter slope IMF\@ 
\citep{salpeter}. 
\citet{hunter95,hunter96} obtained  \filter{F555W} (\filter{V}) and 
\filter{F814W} (\filter{i}) band optical photometry utilizing  HST/WFPC2 in 
order to resolve the cluster's intermediate mass stellar population. 
The IMF  derived for  different annuli out to a radius of 4.7 pc was found to 
be in the range $-1.46 < \Gamma < -1.17$ for the mass range 2.8--15 M$_\odot$, 
again consistent with a Salpeter slope IMF\@. 
\citet{masseyhunter} combined their results for the high--mass IMF with the 
results from \citet{hunter95,hunter96} in order to constrain the IMF from 
2.8 M$_\odot$ up to 120 M$_\odot$. 
Comparing the number of high--mass stars predicted by the intermediate--mass 
IMF from \citet{hunter96}, they found the number of massive stars was 
consistent with a single power--law IMF with a Salpeter slope, 
i.e. $\Gamma=-1.35$.

Combining the two data sets used in \citet{hunter95,hunter96}, \citet{sirianni} 
 derived the IMF between 1.35 M$_\odot$ and 6.5 M$_\odot$, extending the IMF 
determination  into the mass range where the stars are still in their 
pre--main sequence phase. 
The IMF was derived in a box with the dimensions 
$\sim$30\farcs4$\times$26\farcs8\arcsec\ (7.6pc$\times$6.7pc), but 
excluding the inner most 13\farcs6$\times$8.6\arcsec\ (3.5pc$\times$2.2pc). 
Again, a Salpeter slope was found down to  2 M$_\odot$, but the IMF was found 
to be flatter than Salpeter, $\Gamma=-0.27\pm0.08$, between 1.35 M$_\odot$ and 
2 M$_\odot$, suggesting the characteristic mass is higher in this massive, 
metal--poor cluster than $\sim$ 0.5 M$_\odot$ as found  in the Galactic field 
\citep[][]{kro02}.

The foreground (A$_\mathrm{V}=0.7$ mag) and  differential extinction 
(A$_\mathrm{V}\sim0-2$ mag) within the cluster \citep{brandl96} makes 
it desirable to observe the cluster in the infrared, for example the 
\filter{H} band where the extinction is less than 20\% that of the 
\filter{V} band. 
In addition, pre--main sequence stars are often associated with 
circumstellar disks and outflows which will introduce additional 
extinction for the clusters low--mass content. 

We have observed R136 with HST/NICMOS Camera 2 through the \filter{F160W} 
band, which is  similar to a ground--based \filter{H} filter. 
The observations were aimed at being sensitive to  objects  below 1 M$_\odot$ 
for a stellar population with an age of 3 Myr. 
Preliminary results have previously been presented in \citet{HZ1,HZ2}, and 
\citet{MA}. 

The paper is structured as follows. 
The data and their reduction is described in Section 2. 
Section 3 shows the results for the \filter{F160W} band imaging. 
The IMF is derived in Section 4 and compared  with the IMF  derived by 
\citet{sirianni}. 
We point out several plausible reasons for the different results in   
the optical and near--infrared, including mass segregation, and differential 
extinction. 
 Finally, our conclusions  are presented in Section 5.

\section{Data reduction and photometry}
\subsection{Observations}

We have obtained HST/NICMOS Camera 2 images through the \fw\ band  of the 
central 56\arcsec$\times$57\arcsec\ region around R136 in  the  30 Dor 
cluster  (HST program ID 7370).  
The observations were centered on the cluster   
(RA,DEC)=(05:38:43.3,$-$69:06:08) and on two adjacent control fields centered 
on (05:38:42.4,$-$68:52:00), and (05:38:56.9,$-$68:52:00). 
The observing dates were Oct 14 and  16, 1997. 
The  field-of-view of the 256$\times$256 pixel NICMOS Camera 2  is 19\arcsec 
$\times$19\arcsec\  with a pixel scale of 
0\farcs075, resulting in Nyquist sampling of diffraction--limited \fw\ band 
data. 
Each position in a 3$\times$3  mosaic centered on R136 was observed four 
times with small dithers of $\sim$16 pixels.  
The data were obtained in non--destructive MULTIACCUM mode such that the 
photometry of the bright stars can be retrieved due to the first short 
integration in each exposure. 
The integration time for each  dither position was 896 seconds, resulting in a 
total integration time of 3584 seconds for each position in the mosaic. 
The two control fields were observed in a similar manner. 

\begin{figure}
\plotone{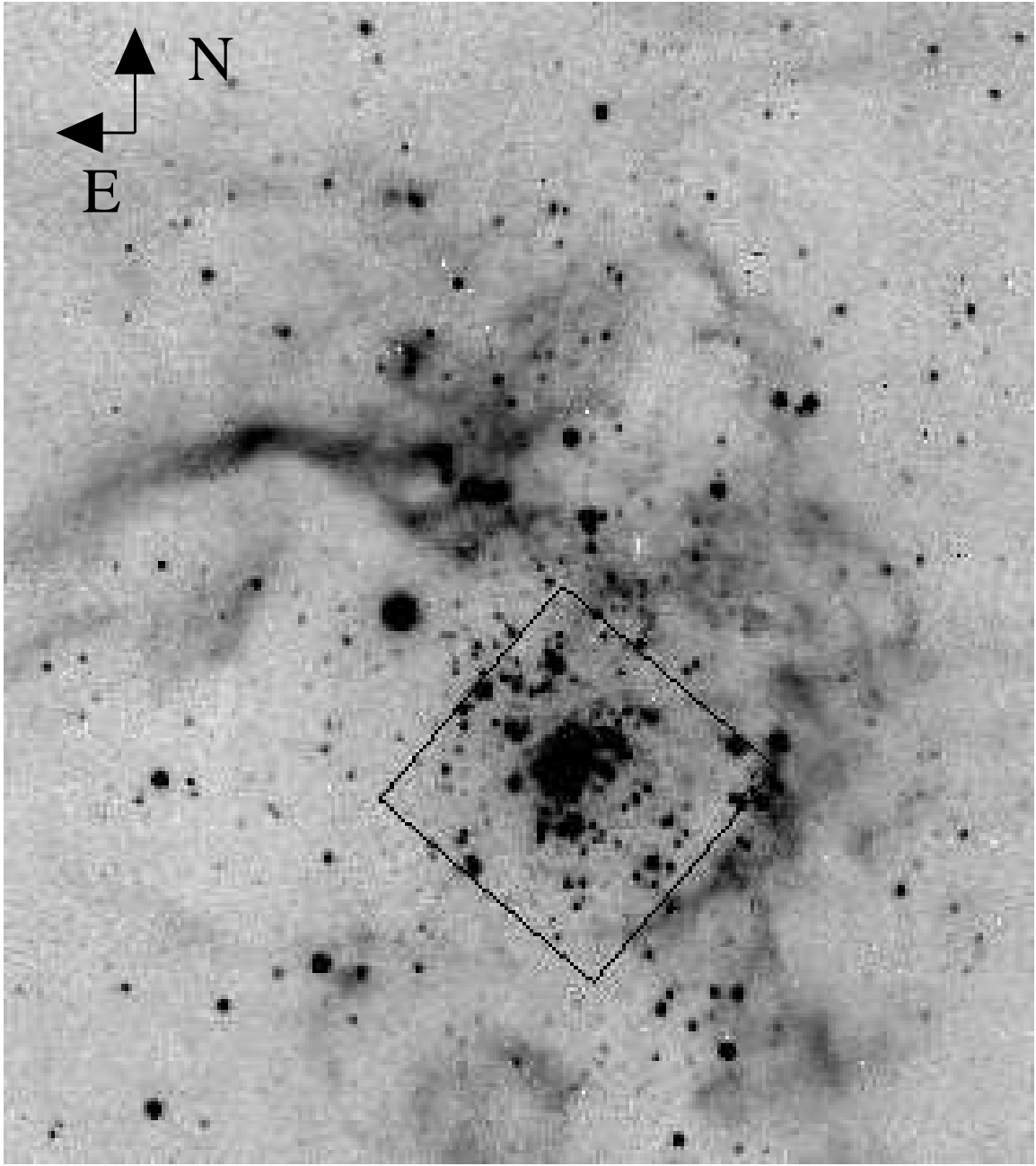}
\caption{Location of the NICMOS surveyed region within  30 Dor, outlined by 
the black box, overlaid on a \filter{K} band image obtained with  the IRAC-2a 
camera on the MPI/ESO 2.2 meter telescope \citep[see][for details]{brandner}. 
The full field is 200\arcsec$\times$225\arcsec and  the NICMOS field is 
56\arcsec$\times$57\arcsec . North is up and East is  left.}
\label{overviewfig}
\end{figure}

\begin{figure*}
\plotone{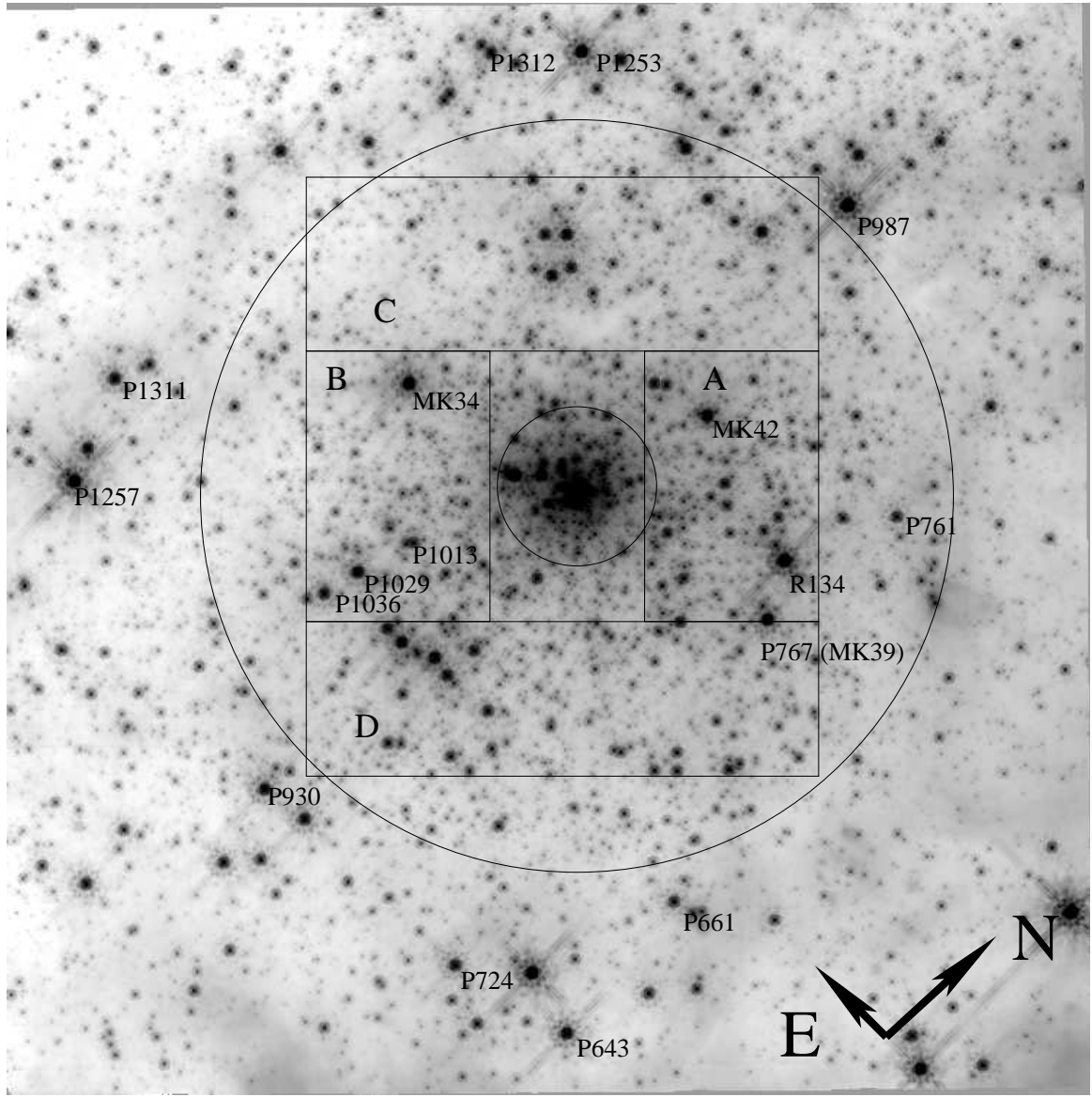}
\caption{The central region around R136 in  the 30 Dor cluster (NGC 2070) as 
observed through the \fw\ filter with the NICMOS Camera 2 on HST. 
The field--of--view is  56\arcsec$\times$57\arcsec, corresponding to 14 pc 
$\times$ 14.25 pc. 
A logarithmic inverted intensity scale has been used. North and East are 
indicated in the Figure. 
The faintest sources visible in the image are \fw\ $\sim$21.5 mag.
Stars with spectral types determined by \citet{parker1} are marked by their 
ID number, along with   R134, MK34, and MK42 \citep{melnick85} to the lower 
right of each star. 
All the identified stars are early O or WR type stars, except P1257 (B0IA), 
P1253 (BN0.5Ia), and P987 (B0.5--0.7I). The spectral types are adopted from 
\citet{WB}. 
The four regions analysed by \citet{sirianni} are shown and labeled. 
Further, the 1 pc and 5 pc radii are indicated by circles. }
\label{30dormos}
\end{figure*}

The location of the mosaic is shown  in Fig.~\ref{overviewfig} and the NICMOS 
mosaic is  shown in Fig.~\ref{30dormos}. 
The faintest stars visible  with the stretch used here  have an \fw\ magnitude 
of $\sim$21.5 mag, corresponding to a mass of 0.8 M$_\odot$, based on the 
pre-main sequence models of \cite{siess}, adopting an age of 3 Myr 
\citep{hunter95}, half solar metallicity, and an extinction of 
A$_\mathrm{V}=1.85$ mag (see Section 3.2). 
For comparison, the similar detection limit in an uncrowded environment without nebulosity would be $\sim$23.5 mag according to the NICMOS exposure time calculator. 

\subsection{Data reduction}
Each individual image was processed through the {\tt calnica} and 
{\tt calnicb} pipelines as well as the {\tt biaseq} and {\tt pedsky} 
procedures within the IRAF environment. 
 The tasks are described in detail in the NICMOS Data Handbook.  
We used synthetic dark frames and flat fields created for the  appropriate 
instrument temperature at each exposure. 
The {\tt biaseq} task  corrects    differences in bias  levels for each chip
between different  sub-exposures. 
The {\tt pedsky} task corrects   differences in the bias level 
for each quadrant of the chip when the array is reset before the exposure. 
The data for each  position in the mosaic were combined  using the  
{\tt drizzle} task.  
The reduced  pixel size   (0\farcs0375) was chosen as half the detector pixel size.
Bad pixels, bad columns, and the coronagraphic  hole were flagged as bad 
pixels before the images were combined. 

\subsection{Source detection and photometry}
Source detection was done using {\tt daofind}  and photometry was performed 
 via point spread function (PSF) photometry utilizing {\tt allstar}  within
 the IRAF environment. 
It was difficult to obtain a good PSF model from the data due to the high 
degree of crowding and the spatial variability of the PSF.  
  Instead, the TINYTIM software \citep{hook} was used  to create a synthetic 
PSF. 
TINYTIM allows to create a PSF that varies as a function of the location on the array. 
The source detection and photometry was performed on each individual position 
in the mosaic due to the linearly varying Point Spread Function (PSF). 
A hot template star (O5V) was used for the spectral energy distribution  in 
order to  achieve the best fit for the brightest stars  to limit their 
residuals. 
The TINYTIM PSF was created for five different positions on the NICMOS 
Camera 2 array and the PSFs were placed in an empty frame with the same 
number of pixels as the NICMOS Camera 2 array. 
Four frames were created with offsets between each PSF identical to the 
offsets used for the science data in order to replicate the data as closely 
as possible. 
The four PSF frames were then combined using drizzle together in the same 
manner as the science data and a linearly--varying PSF  was  created  from 
the drizzled frame.

Source detection is complicated due to the  diffraction features  present in 
NICMOS data.
Adoption of  a low threshold for source detection led to numerous diffraction 
spots from bright stars being  erroneously  identified as  fainter stars. 
Instead  the source detection was done in the following way in order to limit 
false detections. 
We first detected the brightest stars (brighter than 1000$\sigma$) in each 
frame and used {\tt allstar} to remove these with the synthetic PSF. 
 A search for fainter stars (brighter than 500$\sigma$)  was then performed in 
the frame with the bright stars removed. 
Since the removal of the brightest stars also removed the diffraction pattern 
associated with them, we did not detect  the diffraction spots as stars. 
The two star lists (the brightest stars and the fainter stars) were joined 
into one and these stars were removed from the original frame, again using 
{\tt allstar}. 
Fainter stars are then found from the frame with the already detected stars 
removed.
This process was iterated until stars at 10 $\sigma$ peak pixel intensity 
over the background were detected and removed. 
The frame with the stars removed was then ring--median filtered to remove 
stellar residuals but to  retain the large-scale nebulosity in each frame. 
The median--filtered image was then removed from the original frame and the 
star detection process was repeated in this frame but  now continued to a 
detection threshold of 5$\sigma$. 
A 5$\sigma$ instead of e.g. a 3$\sigma$ threshold was selected to limit the 
risks of false detections due to noise spikes. 
We finally made sure by visual inspection that every detection indeed was a 
point source and that it was not a spurious detection due to the diffraction 
spikes and spots from bright stars. 

The main interest here is  in the low--mass (faint) stellar content in R136 
and  one concern is the detection of residuals from the bright stars as false 
stellar objects. 
Some false sources were detected by {\tt daofind} but are rejected during the 
PSF fitting routine. 
A few remained from the brightest stars. 
They  typically produced at most a  few false detections in the diffraction 
spikes that were $\sim$6--7 mag fainter than the bright source. 
We removed these detections together with other false detection through the 
visual inspection of all sources. 
We have further utilized the artificial star experiments described below to 
examine how many detections are false due to the residuals from bright stars. 
We had only false positives associated with the brightest artificial stars  
(\filter{F160W} $<$ 12 mag). 
For artificial stars fainter than \filter{F160W}$\sim14$ mag, no false 
detections were present. 
The false detections for the bright stars were  located at the diffraction 
spikes and would have been identified in the manual inspection of the source 
list. 

We found a total of 10108 uniquely detected sources with a formal error 
smaller than 0.1 mag and brighter than \filter{F160W}=22.5 mag in the 9 frames. 
Below this magnitude limit the incompleteness is substantial, as discussed 
below. 
Table~\ref{sources} presents the list of detected stars. 


\subsection{Completeness corrections}
\begin{figure}
\epsscale{.7}
\plotone{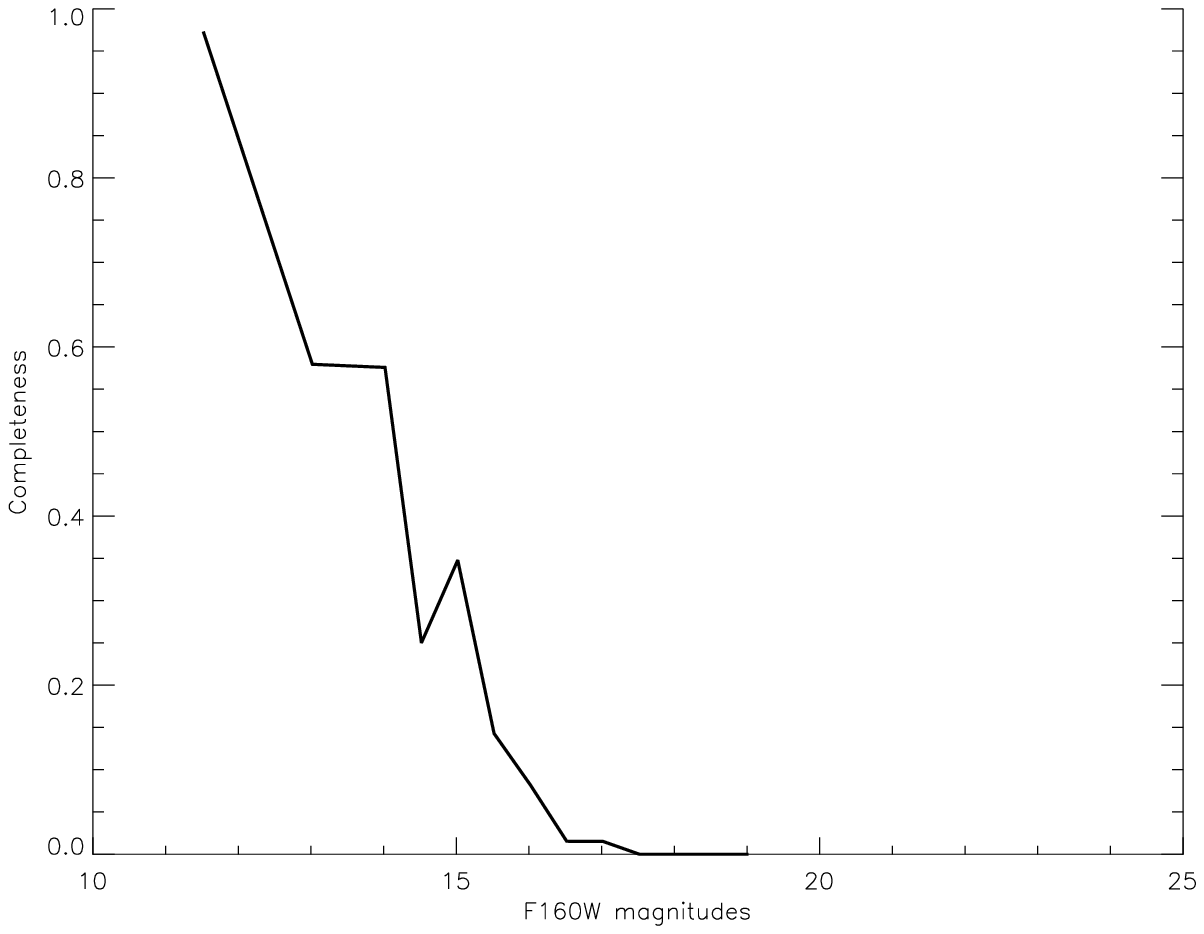}
\caption{The completeness as a function of magnitude for different annuli 
extending from the center to the edge of the field. 
The error bars indicated on every second curve are the statistical 
fluctuations in the completeness tests. 
The corresponding masses for a 3 Myr isochrone and an extinction of 
A$_\mathrm{V}=1.85$ mag are indicated at the top of the plot. 
}
\label{30dor_corrections}
\end{figure}

The effects of crowding were examined by  placing artificial stars in the 
individual frames using the PSF  created from 
the synthetic TINYTIM PSF. 
The artificial stars followed a  luminosity function with a similar slope to 
that of  the observed stars (see Section 3) but with a surface density  10\%\ 
that of the detected number of stars  to avoid affecting the crowding 
characteristics of the real stars. 
We performed 100 artificial star experiments  for each frame, for a total of 
10 times more artificial stars than real stars. 
Fig.~\ref{30dor_corrections} shows the resulting recovery fractions as a 
function of  the input magnitudes for several  annuli around the cluster 
center. 
The difference of the size of the error bars as a function of distance from 
the cluster center is due to a lower number of artificial stars placed in the 
central parts of the cluster. 
This is a consequence of adding  10\% artificial stars relative to observed 
stars  in each artificial star experiment and the relative number of stars in 
each annulus. 
The  IMF is not determined in regions with this low completeness. 
We are mainly interested in the low-mass stellar content of the cluster, which 
is below the 50\%\ completeness in the central parts. 
The uncertainty in the completeness corrections for the inner parts of the 
cluster will therefore not affect the conclusions drawn for the stellar 
populations further out. 

The completeness is a strong function of the radial  distance from the center. 
For the outer regions of the cluster, 50\% or more of the stars brighter than 
\fw$=21.5$ mag are detected, whereas only the very brightest stars are 
detected in the innermost region. 
In an annulus at  0.6-1 pc radius from the center, we detect 50\% or more of 
the stars brighter than \fw$=18.0$ mag. 
Adopting the PMS models of \citet{siess} and the main sequence models of 
\citet{marigo}, \fw$=21.5$ mag corresponds to a 0.8 M$_\odot$, half solar 
metallicity, 3 Myr old object, whereas \fw$=18$ mag corresponds to a  
7.5 M$_\odot$ star, assuming an extinction of A$_\mathrm{V}=1.85$ mag in both 
cases. 

 \subsection{Photometric accuracy}
\begin{figure}
\plottwo{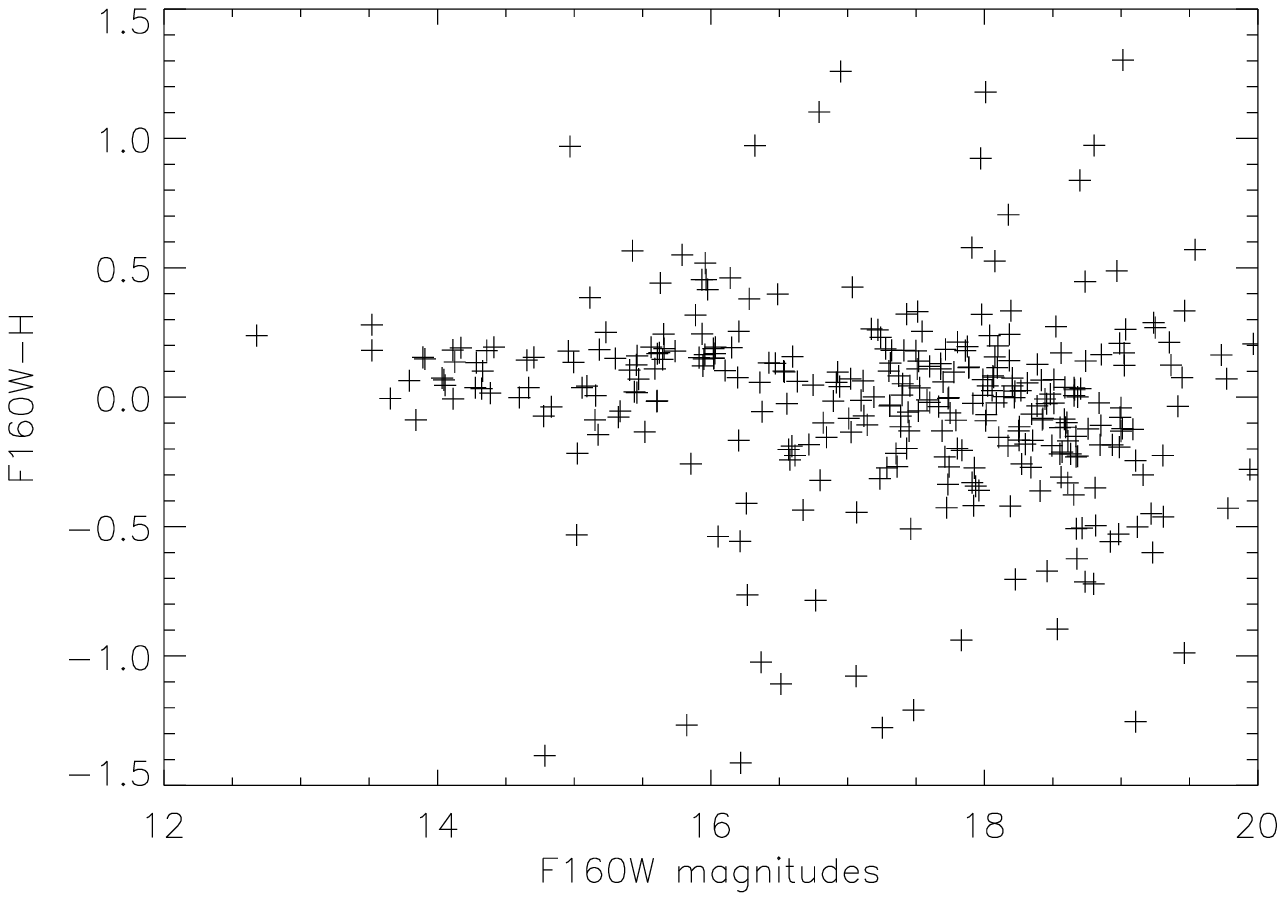}{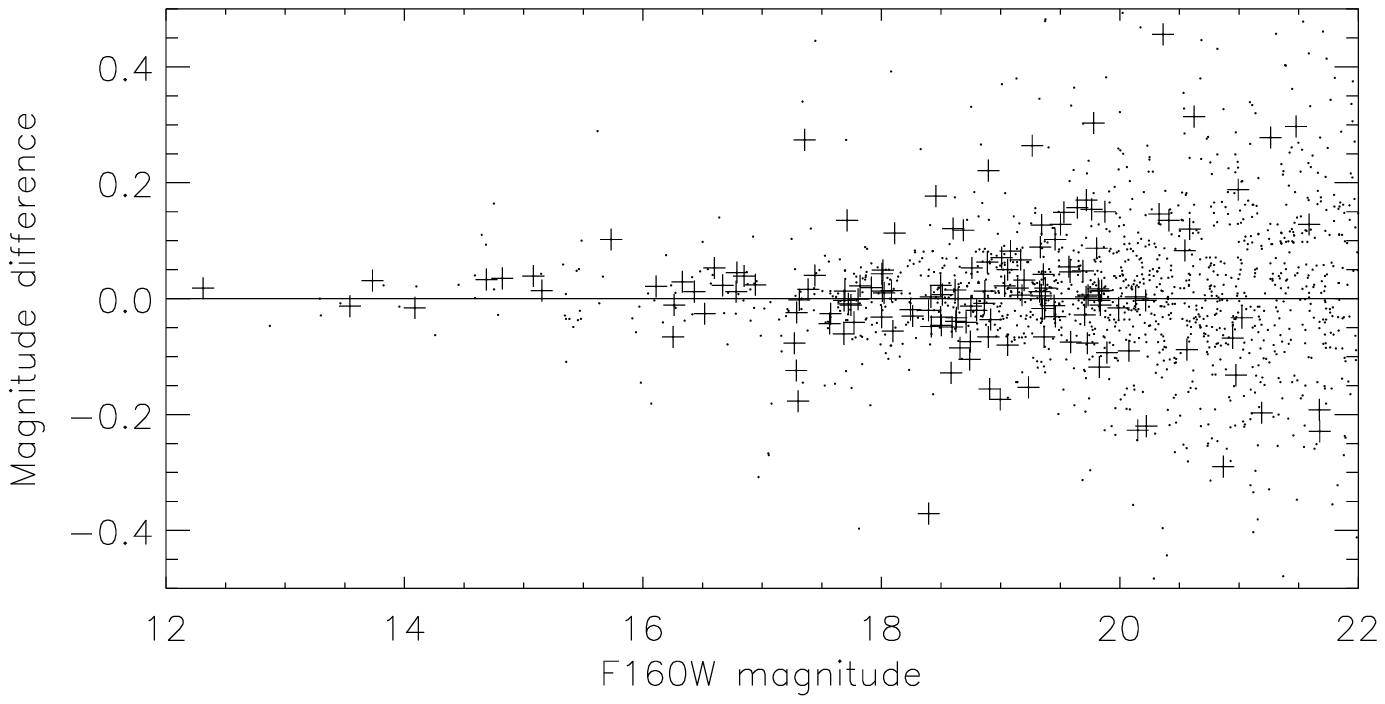}
\caption{Left: A comparison of the NICMOS 2 photometry derived in this study 
with the ground--based AO \filter{H} band  observations by \citet{brandl96}. 
The mean difference between the two datasets for objects with  \filter{F160W} 
$\le$ 17 mag is 0.03 mag, and the standard deviation is 0.2 mag.  
Right: The magnitude difference for sources detected in two different images 
in the mosaic. 
A star was identified in both images if its position was within one pixel. 
Plus signs denote stars within a 2 pc radius of the center, but outside 1.25 
pc; the dots denote stars outside this annulus. 
The RMS in the error for objects with \fw $< 21.5$ mag, which is the 50\%\ 
completeness limit in the outskirts of the cluster,  is 0.17 mag.}
\label{brandl_comp}
\end{figure}

We have investigated the accuracy of the derived photometry  by using the 
stars detected in the overlap regions of several fields. 
Fig.~\ref{brandl_comp}  shows the difference in derived magnitude  for stars 
detected in the overlap regions of the mosaic. 
Dots denote stars outside a 2 pc  radius and plus signs denote stars between 1.25--2 
pc radius, respectively. 




 The \fw\ band photometry has  been compared with the ground--based \h\ band 
photometry obtained using adaptive optics observations by \citet{brandl96}. 
 Fig.~\ref{brandl_comp} shows the magnitude difference between the adaptive 
optics photometry and this study based on 829 stars common to both datasets. 
Stars were considered detected in both datasets if the spatial position 
coincided within 2.5 drizzled pixels, corresponding to 0\farcs094. 
Some scatter is present between the two datasets, especially for the fainter 
stars. 
However, the median difference between the magnitudes derived for the two 
datasets is less than 6\% for objects  \filter{F160W}$< 18$ mag. 
We have in the following treated the \filter{F160W} observations as a 
standard Cousins \filter{H} band.  

Conversely, there appears to be a tendency for the fainter stars to be 
brighter in the \fw\ data than in the \filter{H} band data of \citet{brandl96}.
The tendency for the fainter stars to be skewed towards fainter \h\ band 
magnitudes is an effect also seen in other comparisons between   HST/NICMOS  
and AO data \citep[e.g.][]{stolte} who suggest it is due to  the extended 
halos present in AO observations around bright stars. 

\section{Results}
The immediate results from the \fw\ band HST photometry are presented. 
After discussing the luminosity function for different annuli,  the luminosity 
profile for the cluster is derived. 
The \fw\ band data are combined with the optical HST data by \citet{hunter95} 
and  the color--magnitude diagrams are presented. 
 Utilizing the two color--magnitude diagrams we show that the spread 
observed for the higher mass stars is consistent with that expected due to 
reddening.
We estimate the average age for the stellar population and discuss the 
possible presence of an age spread. 
\subsection{Luminosity functions}
\begin{figure}
\epsscale{.87}
\plotone{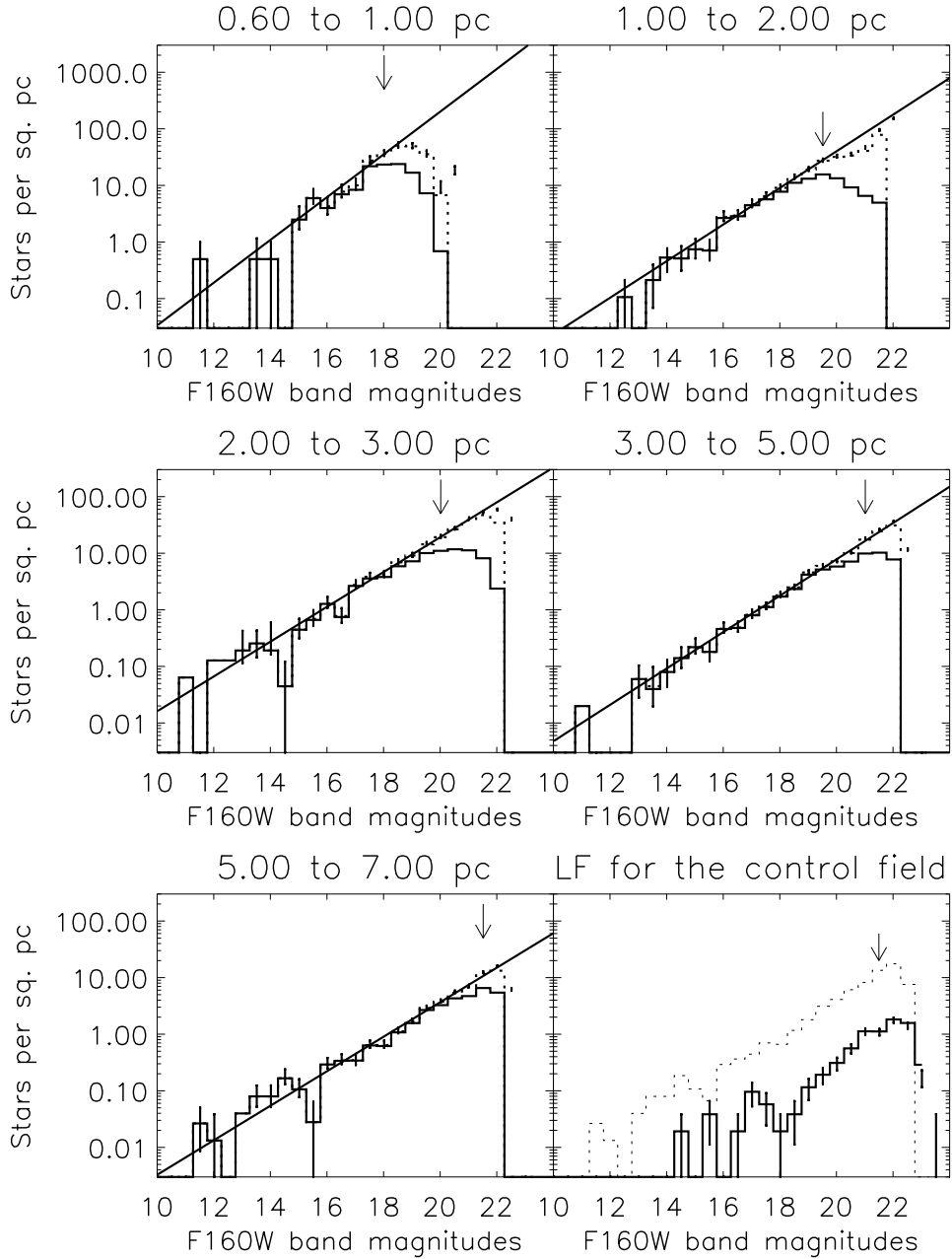}
\caption{The \filter{F160W}--band luminosity functions  for different annuli in 
R136 outside a radius of 0.6 pc. 
Solid histograms indicate the observed number of stars per magnitude bin  
while the completeness--corrected data  are shown as the dotted lined 
histograms. 
The error bars include the Poisson errors and the uncertainty from the 
incompleteness calculations where the two error terms have been added in 
quadrature. 
In each panel, an arrow  indicates the bin where the completeness correction 
is 50\%.  The solid straight line is a weighted fit  to the completeness 
corrected histograms down to the 50\%\ completeness limit. 
The lower right panel shows the completeness corrected average luminosity 
function for the two control fields  (solid histogram) associated with the 
30 Dor cluster. 
The completeness corrected luminosity function for the 5--7 pc annulus in 
30 Dor is shown as the dashed histogram for comparison. }
\label{LFs}
\end{figure}

The star counts in the central 0.6 pc radius region is heavily affected by low 
number statistics, crowding even for the brightest stars,  and relatively 
uncertain incompleteness corrections. 
We therefore focus on the sources outside 0.6 pc in this paper. 
 Fig.~\ref{LFs} shows the \fw\ band luminosity functions for the 0.6--7 pc 
radius region of the 30 Dor cluster divided into several radial bins to show 
the difference in photometric depth due to crowding. 
Overplotted are the completeness--corrected LFs, where each bin has been 
divided by the corresponding  recovery fraction from the artificial star 
experiments.  

The completeness--corrected luminosity functions are relatively smooth and  
have been  fitted with power-laws down to the 50\% completeness limit. 
The derived slopes with their 1$\sigma$ uncertainties and the 50\%\ 
completeness limits are presented in Table~\ref{HLF_slopes}. 
%
Although the slope in the inner annulus is found to be more shallow, the derived slopes are consistent with each other within 2$\sigma$ with an average slope 
of 0.31 and the shallow slope is not significant.

The completeness--corrected combined histogram for the stars detected in the 
two off--cluster control fields is shown in the lower right panel in 
Fig.~\ref{LFs}. 
From the histogram, it  can be seen that the  field star contamination found 
from the star counts is $\approx$10--15\%\ for the faintest stars in the 5--7 
pc annulus and less closer to the center as well as  for  brighter stars. 
Although the contamination of field stars is found to be relatively small  it 
is not negligible and  they are therefore   statistically subtracted from the 
cluster population in the following analysis (Section 4). 

\subsection{The optical--near infrared color--magnitude diagrams}
Next, the \fw\ band photometry is combined with the optical data presented by 
\citet{hunter95}. 
A star was considered detected in both surveys if the spatial position agreed 
within 2.5 drizzled NICMOS Camera 2 pixels (0\farcs094). 
In the cases where two optical stars were located within the search radius of 
the star detected in the NICMOS Camera 2 observations, the brightest star was 
chosen as the match.

We find in total 2680 in common with the \citet{hunter95}  survey that 
detected 3623 stars the inner 35\arcsec\ of the cluster. 
1848 of those sources have a combined formal  photometric error in the 
\filter{F555W}--\filter{F160W} color of less than 0.1 mag. 
Within the area covered by \citet{hunter95} we detect a total of 5095 sources. 
Most of the stars detected by the NICMOS survey but not the WFPC2 observations 
are fainter than \filter{F160W}$=$ 20 mag. 
Assuming an object age of 3 Myr and an average  extinction of 
A$_\mathrm{V}=1.85$ mag (see below), the similar object would have a magnitude 
in the \filter{F814W} band of $\sim$22 mag. 
For objects with more extinction, they will be even harder to detect in the 
\filter{F814W} band. 
\citet{hunter95}  essentially don't detect   any stars within a 1 pc radius at 
this magnitude or fainter. 
 Only 1 in 4 stars in the magnitude interval \filter{F814W}=21--22 mag was 
detected  outside 1 pc. 
It is thus not surprising that a significant population of faint stars are 
detected in the NICMOS survey relative to the WFPC2 survey. 
Nevertheless, the lower spatial resolution of the NICMOS observations results 
in a low recovery fraction at these magnitudes in the central few pc of the 
cluster. 

The majority of the sources not detected in the NICMOS survey but at optical 
wavelengths are located within a radius of 1 pc. 
The lack of detection is due to the lower spatial resolution in this study 
relative to the optical HST data.  
The resolution is almost a factor of two better in the \filter{F814W} band 
than in the \filter{F160W} band. 
Sources not detected in the NICMOS data outside 1 pc are mainly due to 
crowding as well. 
Indeed, visual inspection of the location of the stars detected in the optical 
but not near--infrared shows they are often located either very close to the core 
or on the first Airy ring of a bright source.

\begin{figure}[t]
\begin{center}
\epsscale{1.1}
\plotone{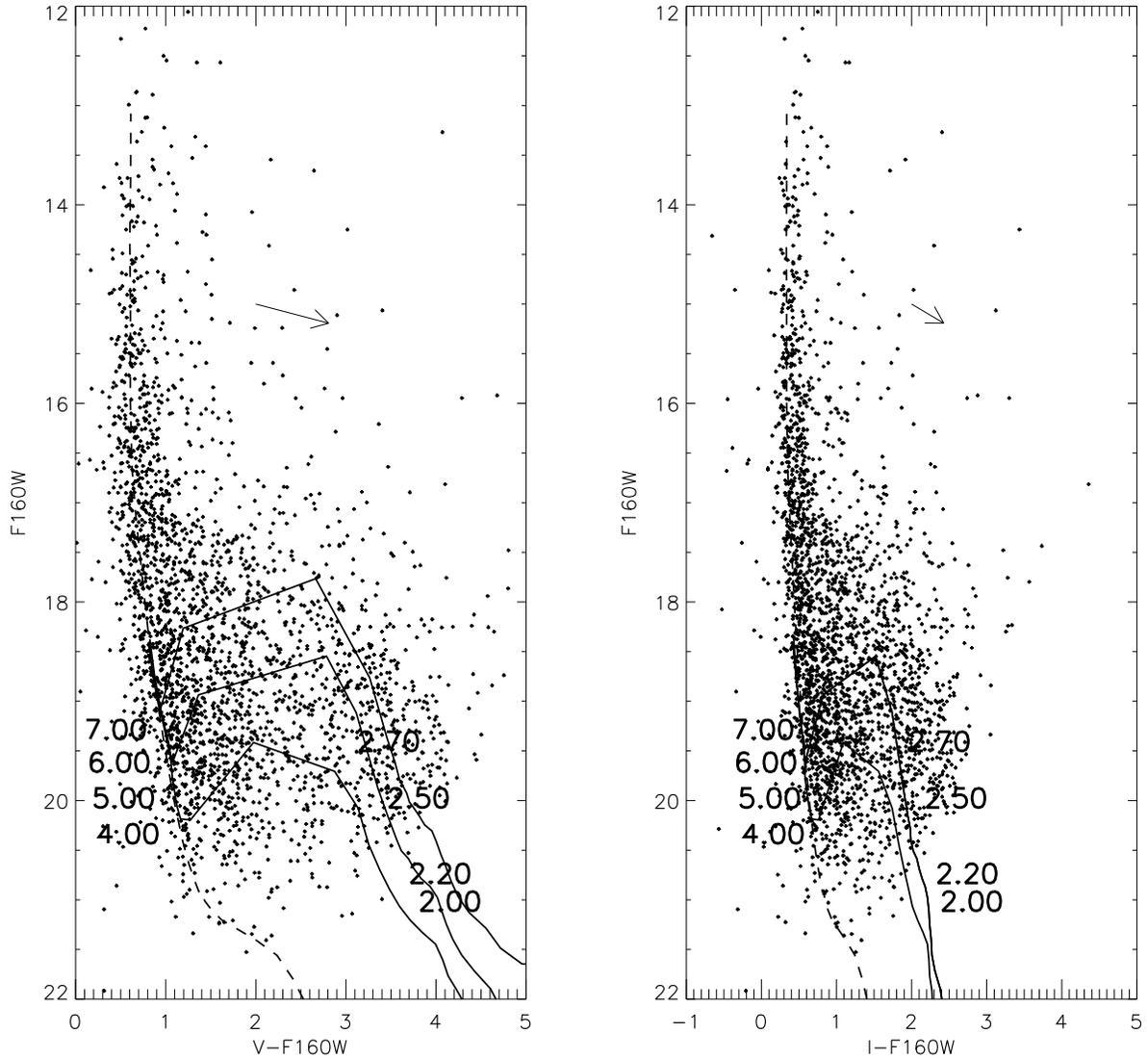}
\caption{Left: The \filter{F555W}--\fw\ versus \filter{F160W} 
color--magnitude diagram for R136 for the central 35\arcsec$\times$35\arcsec\ 
observed by \citet{hunter95} in the \filter{F555W} and \filter{F814W} bands. 
The arrow  illustrates the effect of A$_\mathrm{V}=1$ mag of extinction 
(A$_\mathrm{F160W}=0.175\times$A$_\mathrm{V}$). 
Overplotted as the solid lines are the 2, 3, and 4 Myr pre--main sequence 
isochrones from \citet{siess} below 7 M$_\odot$. 
The long dashed line is the main sequence from \citet{marigo} below 
7 M$_\odot$  and the  3 Myr isochrone from a \citet{marigo} above 7 M$_\odot$. 
All the  isochrones have been reddened by A$_\mathrm{V}=1.85$ mag, the median 
reddening for the high mass stars (see the text) and shifted by a distance 
modulus of 18.5. 
The  \filter{F555W}  magnitudes have been transformed into the \filter{V} 
magnitude using the transformations in  \citet{holtzman}. 
The 3 Myr \citet{siess} isochrone  joins the main sequence at 
\filter{F160W}$\sim$19.5 mag, \filter{V}--\fw$=$0.8 mag. 
Masses are indicated on the 2 Myr isochrone. 
Right: Same as to the left but this time the I--\fw versus \fw\ 
color--magnitude diagram. 
Notice the tighter clustering  around the isochrone to the right, 
indicating part of the scatter in the V-\fw\ color--magnitude diagram is 
due to differential reddening. }
\label{CMD}
\end{center}
\end{figure}

The \filter{F555W}-\fw\ versus \filter{F160W} color--magnitude diagram is 
shown in Fig.~\ref{CMD}. 
Overplotted are a  3 Myr isochrone  for the high--mass stars adopted from the 
\citet{marigo} models and   2, 3, and 4 Myr isochrones adopted from 
\citet{siess} for stars below 7 M$_\odot$. 
The stars above 7 M$_\odot$ and up to the maximum mass we fit the IMF in 
Section 4.1 (20 M$_\odot$)  are all expected to be on the main sequence. 
Both isochrones were calculated adopting a metallicity of half the solar 
value, typical for the LMC \citep{smith}. 
The two isochrones have a small offset in both the \filter{V} (0.06 mag) and 
\filter{H} (0.07 mag) band. 
We have forced the \citet{marigo} isochrone to match the \citet{siess} 
isochrone at 7 M$_\odot$.

It is evident there is a significant scatter in the color--magnitude diagram. 
The scatter is likely due to a combination of binary systems (both physical 
and chance alignments), differential extinction, photometric errors and a 
possible  age spread. 
The median extinction is found for the main sequence part of the isochrone. 
For objects in the range 7--20 M$_\odot$, we find a median extinction of 
A$_\mathrm{V}=1.85$ mag which is slightly higher than   the reddening found by 
\citet{selman2} in the inner part of the 30 Dor region. 

At masses below 7 M$_\odot$, the spread in the color--magnitude diagram is 
larger but almost exclusively extends to the red part of the diagram. 
This indicates the lower--mass objects on average have an excess amount of 
extinction relative to the higher mass objects. 
\citet{selman2} observed stars more massive than $10$ M$_\odot$ and would not 
detect the additional reddening for the lower--mass sources. 
The possible sources for  the additional reddening is described in Sec.~3.3.

We have estimated  an average age for the cluster by utilising the fact the 
isochrone is almost horizontal in the color range 
\filter{F555W}-\filter{F160W}=1.5--2.5 mag and around 
\filter{F160W}$\sim$19 mag. 
The  median  \filter{F160W} magnitude is 19.0 mag in this region of the 
color--magnitude diagram. 
Adopting an average extinction of A$_\mathrm{V}=1.85$ mag,  this corresponds  
to the \filter{F160W} magnitudes of the 3 Myr isochrone in the same 
color range. 
We have thus adopted 3 Myr as the mean age of the low mass cluster population 
and a 3 Myr isochrone is adopted to create a mass--luminosity relation in 
order to turn the luminosities into masses for objects below 7 M$_\odot$ and 
the 3 Myr \citet{marigo} isochrone above. 
We will in Sect.~4 the effects on the derived IMF  adopting an age spread 
of 2 Myr. 

 The right hand panel in Fig.~\ref{CMD} shows the \filter{I}--\fw\ versus 
\fw\ color--magnitude diagram. 
It is evident that the clustering around the isochrone is tighter than for 
the \filter{V}--\fw\ versus \fw\ color magnitude diagram. 
This is expected if a large part of the scatter is due to differential 
extinction. 
We can calculate the scatter around the main sequence in both color--magnitude 
diagrams and compare with the difference predicted from extinction. 

If the spread in the color--magnitude diagrams is due to extinction we expect 
the ratio of spread in the \filter{V}--\fw\ versus \fw\ diagram to be ratio of 
the extinction in each color, i.e. $(1-0.192)/(0.62-0.192)=1.88$ times larger 
than in the \filter{I}--\fw\ versus \fw\ color--magnitude diagram. 
Since the isochrone is almost vertical in both diagrams, we have calculated 
the standard deviation around the reddened isochrone in both color--magnitude 
diagrams. 
We have used the stars with good photometry, better than 5\%\ in each filter, 
and in the magnitude range $13 < \fw < 17$mag. 
The standard deviation found for the \filter{V}--\fw\ and \filter{I}--\fw\ 
color--magnitude diagrams are 0.60 mag, and 0.36 mag and the ratio is 1.7. 
If the measurement errors are taken into account this ratio increases. 
The typical errors for the culled sample are 0.04, 0.02, and 0.03 mag for the 
V, I, and \fw\ bands, respectively. 
After taking the measurement errors into account, the  ratio is found to be 
1.9, assuming the measurement errors in two filters are independent. 
Due to blending, this is not necessarily the case. 
Thus,  1.9 is an upper limit and we thus find the ratio to be between 
1.7 and 1.9, in agreement with the scatter being due to differential 
extinction. 

Since the amount of differential extinction does not affect the \fw\ band 
photometry significantly, the single band photometry presented here is 
competitive with the 2--band optical photometry. 
There is a unique translation from the \fw\ band magnitude to the object 
mass for the majority of the mass range. 
For the optical photometry, the color information is used to determine the 
extinction and the mass function is thus effectively determined by the 
de--reddened \filter{V} band magnitude.


\subsection{The differences between the optical and near--infrared HST observations}
The main advantage of the optical relative to the near--infrared HST 
photometry is the improved resolution due to the smaller diffraction limit. 
The stellar content  can therefore be resolved to lower masses closer to 
the cluster core than is possible with the near--infrared observations. 
However, phenomena associated with the star formation process can introduce 
additional reddening that can complicate the derivation of the low--mass IMF 
from optical data. 
The low--mass objects may still be associated with a circumstellar disk. 
There is evidence from e.g. the Orion Nebula Cluster that  circumstellar 
disks can survive the UV radiation from massive stars \citep{robberto}. 
Even if the disks are being evaporated by the ration field from the early 
type stars, the evaporated material will be a further source of reddening. 
Patchy extinction associated with the 30 Dor complex and located in the 
foreground of R 136 will be an additional  source of differential reddening. 
There are signs in the optical images presented in Fig. 1 of \citet{sirianni} 
of patches of extinction, e.g. to the east--north--east of the cluster center. 
If  variable extinction is present or if a significant fraction of the stars 
are associated with disks or outflows, an extinction limited sample  has to 
be created in order to avoid  a biases against detection of the 
low--mass stars. 

The near--infrared photometry is effected by differential extinction as well 
but the effect is less than 20\% of that measured in the \filter{V} band.  
Thus, whereas the IMF derived from optical observations where an extinction 
limited sample is not defined might be severely affected for the low--mass 
objects, the effect on near--infrared observations is modest. 
Therefore, in the outer parts of the cluster where crowding is a smaller 
issue than closer to the center, the near--infrared observations are more 
suitable to detect and characterise the low--mass stellar population in the 
cluster.

On the other hand, single band  photometry has the disadvantage that there 
is no information on the age of  individual objects. 
We investigate in the next Section how this might affect the derived IMF. 
We note that if differential extinction is present, the situation is no 
better  for the optical photometry. 
Even though the cluster was observed through two filters in the optical,
 there is still a degeneracy between age and extinction. 
\cite{sirianni} converted the \filter{V}--\filter{i} photometry into an 
effective temperature and used that effective temperature to obtain a 
bolometric correction. 
Without de--reddening the sources, the age of a cluster member  can be in 
error and hence the mass estimates will be uncertain.

\section{Analysis} 

We construct a mass--luminosity relation by combining the main sequence  
models by \citet{marigo}, and the pre--main sequence models of  \citet{siess} 
in order to infer the stellar mass  from the \filter{F160W} band magnitude. 
We then derive the mass functions for R136 outside 0.6 pc  where the 50\% 
completeness limit corresponds to a stellar mass below 10 M$_\odot$. 
Deriving the IMF this way is a well established procedure 
\citep[][]{lada,muench02}. 
We  further discuss the potential effect of extinction on the derived IMF\@. 
Finally, we search for evidence for mass  segregation in the outer parts of 
the cluster using  the cumulative luminosity functions. 

\subsection{Deriving the mass function}
\begin{figure}
\begin{center}
\epsscale{.8}
\plotone{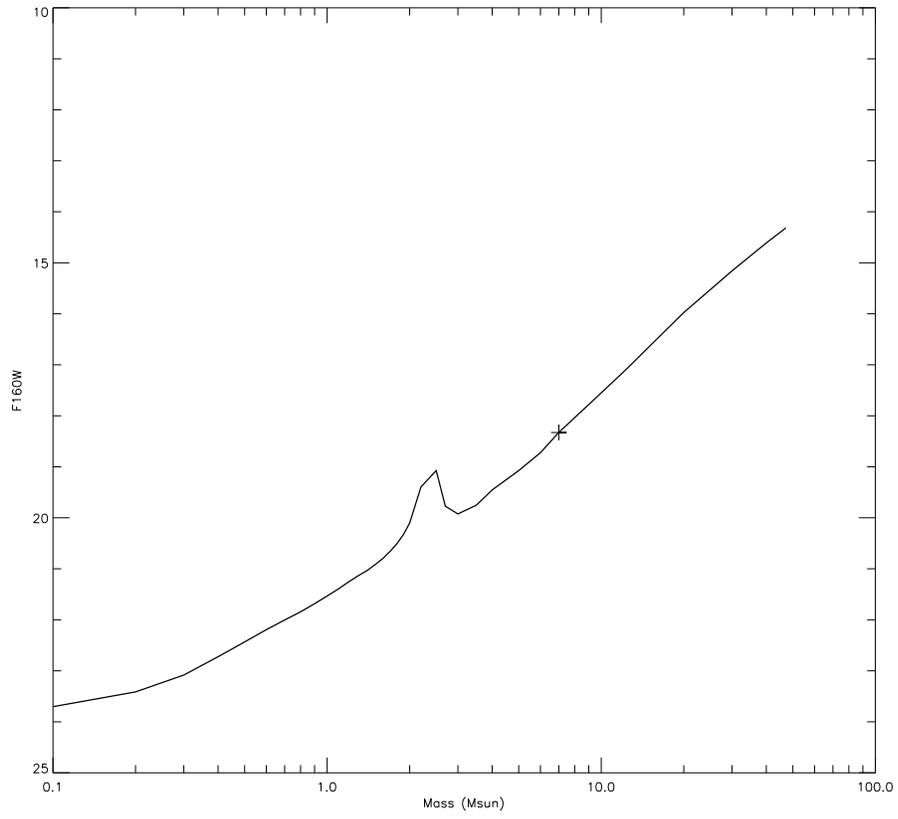}

\caption{The mass--luminosity relation for a 3 Myr isochrone. 
Marked is the location where the \citet{siess} and \citet{marigo} isochrones 
merge (7 M$_\odot$). 
The feature at 2--3 M$_\odot$ is due to the pre--main sequence to main sequence transition. }
\label{MLrel}
\end{center}
\end{figure}

A mass--luminosity relation is needed to  convert the derived \filter{F160W} 
band magnitude for each star to a mass. 
We use the \citet{siess}  isochrones for stars below 7 M$_\odot$ and the 
\citet{marigo} 3 Myr isochrone for the more massive stars as discussed in 
Section 3.2. 
The age of the cluster is first assumed to be 3 Myr and is later varied to 
examine the effects on the derived IMF for different cluster ages. 
Stars below $\sim$3 M$_\odot$ are on the pre--main sequence isochrone whereas 
the more massive stars up to our upper mass limit of 20 M$_\odot$ (see below) 
are on the main sequence.  
The adopted mass--luminosity relation is shown in Fig.~\ref{MLrel}. 


We have limited knowledge of the extinction for the majority of our  objects. 
Instead, we have adopted an average extinction of A$_\mathrm{V}=1.85$ mag, as 
determined from the \filter{V-H} versus \filter{H} color--magnitude diagram 
in Fig.~\ref{CMD}.  
Since the amount of extinction ranges between A$_\mathrm{V}=0.7-3$ mag 
\citep{brandl96} the extinction for an individual object might be wrong by up 
to A$_\mathrm{V}\sim 1$ mag, a maximum error $<$  0.2 mag in the \fw\ band. 
This corresponds to an error of $\sim$10\% when the luminosity is transformed 
into a mass. 

\begin{figure}[t]
\begin{center}
\epsscale{.8}
\plotone{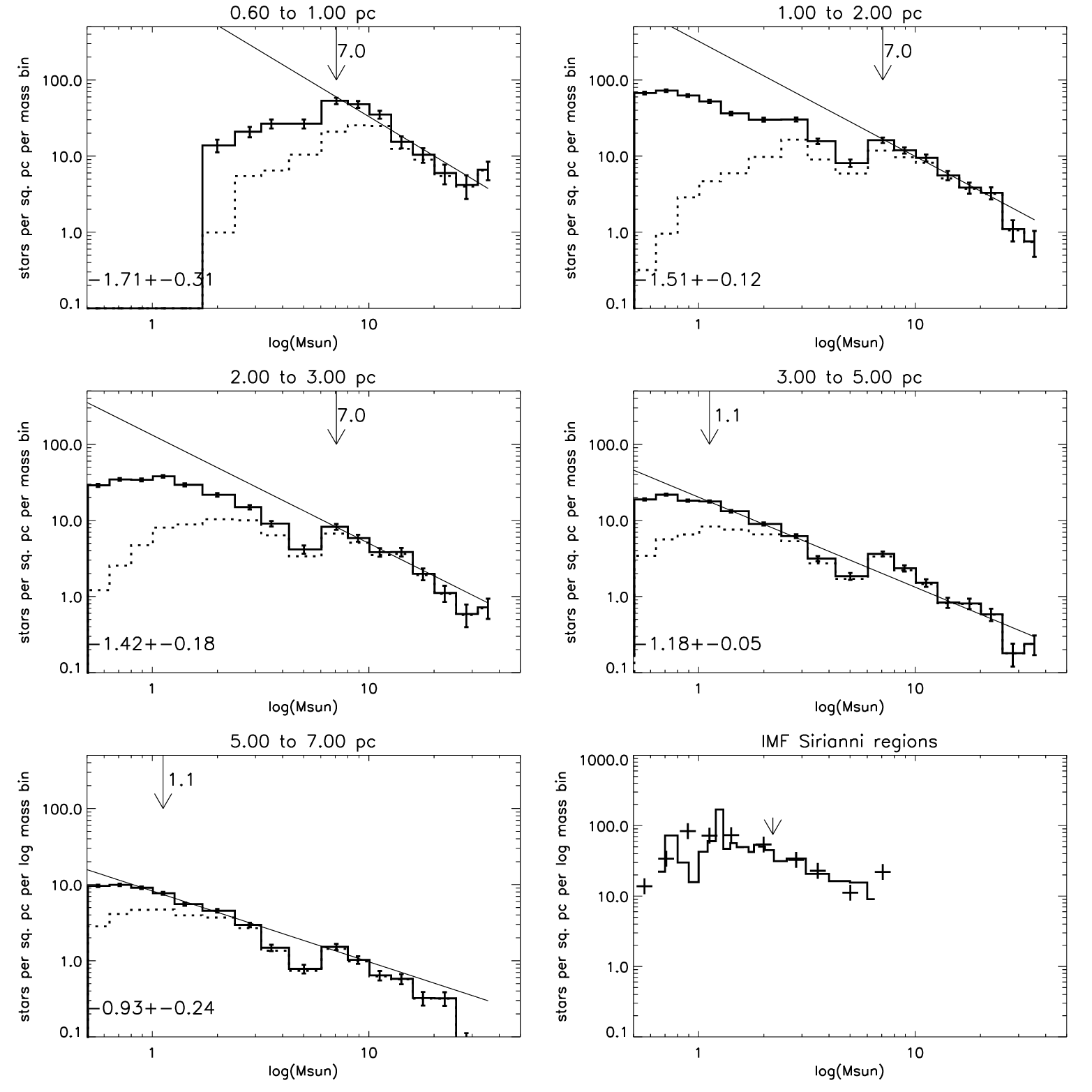}
\caption{The mass functions for 30 Dor outside 0.6 pc in several annuli after 
field star subtraction. 
The arrows indicate the 50\%\ completeness limit. 
Dotted line histograms show the mass functions derived from the uncorrected 
star counts whereas the solid line histograms are the completeness corrected 
mass function. 
Expected errors due to Poisson noise  are indicated on the solid line 
histograms.  
A fit is made to the completeness corrected histogram, corrected for field 
star contamination.  
The maximum mass used in the fits is 20 M$_\odot$. 
The coefficient shown in  each panel is $\Gamma$, $dN/d\log M\propto M^{\Gamma}$. 
The lower right panel shows the IMF derived by \citet{sirianni} from  the 
areas shown in Fig.~\ref{30dormos}. 
Further, the plus symbols show the IMF derived from the NICMOS data from 
the same regions as was used in the \citet{sirianni} study. 
The 50\% completeness is 2 M$_\odot$ for this sample, as shown by the arrow. }
\label{massfuncs}
\end{center}
\end{figure}

Fig.~\ref{massfuncs} shows the derived mass functions outside 0.6 pc for a 
3 Myr isochrone  after field stars have been subtracted statistically in 
each annulus. 
The mass functions are in general smooth and well fit by power--laws. 
However, there appears to be some structure in the derived IMFs at 
intermediate masses, 2--4 M$_\odot$, which is the region where the pre--main 
sequence track joins the main sequence. 
The mass--luminosity relation is plagued by  a non--monotonous feature at 
this mass range (see Fig.~\ref{MLrel}), which marks the radiative-convective 
gap \citep{mayne} 
and the transition region from 
pre-main sequence to main sequence \citep[see also][]{stolte04}. 
A similar structure in the derived IMF is seen in the results from e.g. 
NGC 3603 but at a slightly higher mass since the cluster is 
younger \citep[e.g.][]{stolte2}. 
The turn--on mass is higher for a younger cluster.
Thus we would expect the kink in the mass--luminosity relation to move to 
higher masses for a younger cluster. 
Since this is what is seen comparing NGC 3603 and R 136,  it indicates 
indeed a feature of the isochrones and not a feature intrinsic to the cluster. 
The number of stars in each mass bin is provided in Table~\ref{numbers} .

Power--laws have been fitted to each of the histograms in order to derive 
the slopes of the mass function in each annulus. 
The fit was done over the mass range from 20 M$_\odot$ down to the 50\%\ 
completeness limit for each annulus. 
The mass for stars above $\sim$20 M$_\odot$ is very poorly constrained from 
near--infrared observations  due to uncertainties in the bolometric 
corrections \citep[e.g.][]{massey03}. 
The derived slopes $\Gamma$, where $dN/d\log M\propto M^{\Gamma}$,  are 
indicated in Fig.~\ref{massfuncs} and are also presented in Table~\ref{slopes}. 
The derived slopes for annuli outside 1 pc are consistent with each other 
within 2$\sigma$ error bars.  
For the 3--5 pc and 5--7 pc  annuli  where the data are complete to below 
2 M$_\odot$, the slopes are found to be $-1.2\pm0.1$ and $-0.9\pm0.2$, respectively,  slightly shallower than the slope of $\Gamma=$-$1.28\pm0.05$ derived by 
\citet{sirianni} above 2 M$_\odot$, except that in our case the IMF continues 
as a power--law down to 0.8 M$_\odot$. 


Has the fact that we used the whole mass range for our power-law fit
 washed out a possible 
flattening at the low mass end?  To test this possibility, we have 
additionally fitted a separate power--law to the low--mass part of the IMF. 
Only the part of the mass function that is not influenced by the kink in the 
mass luminosity relation is used. 
This region is limited to masses below 1.7 M$_\odot$ for the 3 Myr isochrone. 
It is therefore only for the 5--7 pc annulus that a reasonable mass range is 
covered to fit the IMF. 
We find the slope to be $\Gamma=-0.9\pm0.2$, which is more shallow but 
consistent at the 2$\sigma$ level with a Salpeter IMF and is consistent 
with the slope derived for the full mass range.  

We have derived the IMF in the same boxes as done by \citet{sirianni}. 
The completeness correction was calculated independently for each box before 
the IMFs were combined to the average IMF for direct comparison with the 
IMF presented by \citet{sirianni}. 
The 50\% completeness limit for the NICMOS data varies from 2.8 to 
1.4 M$_\odot$ for the four boxes. 
Following Sirianni et al., we have derived an average completeness limit 
for the three regions of 2.2 M$_\odot$. 
As evident, the agreement is good for the common mass range. 
We appear to underestimate the stars at $\sim$ 6M$_\odot$ compared to 
\citet{sirianni}. However those appear to be  recovered at 8 M$_\odot$. 



\begin{figure}
\plotone{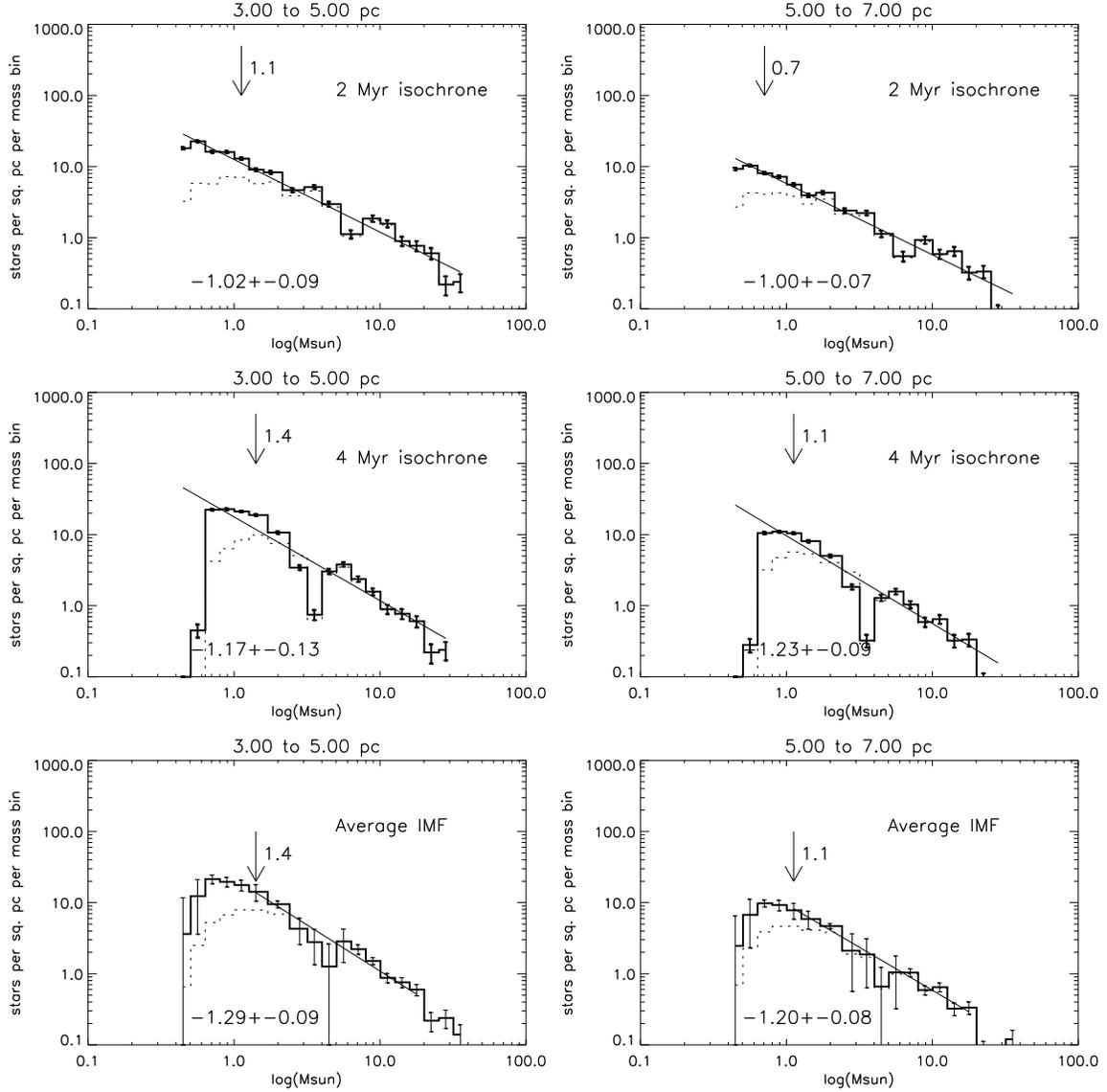}
\caption{The derived IMF's between 3--7 pc assuming a 2 Myr isochrone (top), 
a 4 Myr isochrone (middle) and an average of the IMF's derived for the  2 to 
4 Myr isochrones in 0.5 Myr increments (bottom).
The error bars on the top two panels include  Poisson noise and the 
uncertainties in the completeness corrections. 
The number of stars in each mass bin in the average IMF is determined as 
the average of the 5 IMFs for ages 2--4 Myr. 
The error bars for each bin in the average IMF are then  calculated as the 
standard deviation of the number of stars in the bin.  
The arrow shows the 50\% completeness limit in each panel and the number next to arrow indicate the limiting mass in solar masses. 
For the average IMF, the 50\% completeness limit is determined by the 4 Myr 
isochrone since here the completeness limit corresponds to the highest mass. 
}

\label{IMF_spread}
\end{figure}

The color--magnitude diagrams show a large spread in the main sequence to pre--main 
sequence transition at \filter{F160W}=18--19 mag. 
Although shown in Section 3.2 that this scatter can be explained by differential 
extinction, it cannot be ruled out that there is an age spread present as well as 
suggested in previous studies \citep{hunter95,masseyhunter}. 
It is therefore reasonable to take a  star formation history different than a 
single burst at 3 Myr into account. 
We show in Fig.~\ref{IMF_spread} the IMF in the outer two annuli assuming a 
cluster age of 2 and 4 Myr, respectively. 
We also show an 'average' IMF found as the average of the IMF's derived for 
the age range 2--4 Myr in 0.5 Myr increments. 
The lower mass limit in the average IMF was determined from the 4 Myr 
isochrone which provides the most restrictive mass limit. 
We find that both in the case of a 2 and 4 Myr isochrone the IMF is well fit 
by  power--laws. 
The derived slopes are steeper assuming an older isochrone relative to the 
younger ones. 
There is no indication for a flattening below 2 M$_\odot$ in either case. 
The average IMF is also found to be  represented by a power--law with a 
slope consistent with a Salpeter slope. 
The slopes of the derived power--laws are given in Table~\ref{slopes}. 
For the average IMF,  the number of stars averaged over the different ages 
in each mass bin is derived. 
 Error bars for the average IMF have been determined as the standard 
deviation around the mean number of objects in each mass bin. 

As was the case for the 3 Myr isochrone, the slopes of the IMF for 
different assumed ages have also been calculated and are provided in 
Table~\ref{slopes}. 
The slopes are found to be shallower than a Salpeter slope, 
but at the $\sim2\sigma$ 
level consistent with  a Salpeter slope. 
The slopes are also consistent with those derived for all masses up to 
20 M$_\odot$, as was the case assuming the 3 Myr isochrone.





The lack of a flattening in the IMF below 2 M$_\odot$ is  in contrast to the 
results presented by \citet{sirianni}, who derived the IMF closer to the 
cluster center. 
There can be  several possible reasons for the difference in the derived 
IMF slope in the two surveys. 
First,  due to the different spatial resolution in the two studies, the 
NICMOS IMF is derived  further away from the center of the cluster than the 
WFPC2 IMF by \citet{sirianni}. 
The IMF was derived in the areas shown in Fig.~\ref{30dormos} as regions B,C, 
and D. 
Thus, all of   their surveyed area is outside a radius of 1 pc and the 
majority of their surveyed area is between  2 and 5 pc where crowding precludes 
NICMOS 
from detecting stars less massive than 2.2 M$_\odot$ for a 3 Myr isochrone. 
One possibility for the difference in the derived slopes for the NICMOS and 
WFPC2 data can therefore be   a variation of the IMF as a function of radius. 
Another possible reason  can be  differential extinction as suggested by 
\citet{selman2}. 
Both possibilities  are  discussed in subsections 4.2 and 4.3

\subsection{The effect of differential extinction}
As was suggested by \citet{selman2}, the presence of differential extinction 
can potentially alter the low--mass end of the IMF if an extinction-limited 
sample is not used. 
In order to estimate the possible effect on the IMF if differential extinction 
is not taken into account, we have constructed a simple model of the cluster 
which includes    differential extinction and the depth of the 
dataset from \citet{sirianni}. 
A Salpeter slope IMF and a cluster age of 3 Myr were assumed. 
Each object within the artificial cluster was  assigned a \filter{V} 
band magnitude based on its mass from the 3 Myr isochrone computed for 
a half solar metallicity by \citet{siess}. 
The objects were then reddened by a foreground extinction chosen randomly 
from a normal distribution with a standard deviation of 0.7 and shifted 
to peak at A$_\mathrm{V}$=1.85 mag. 
If the extinction was found to be less than A$_\mathrm{V}=0.7$ mag, 
a new extinction was calculated. 
Stars were then considered detected if their reddened magnitude is within 
the 50\%\ completeness limit presented by \citet{sirianni}. 
The model is obviously an oversimplification of the real situation. 
Nevertheless it  is expected to illustrate how the derived IMF might differ 
from the underlying IMF. 

\begin{figure}[t]
\begin{center}
\plotone{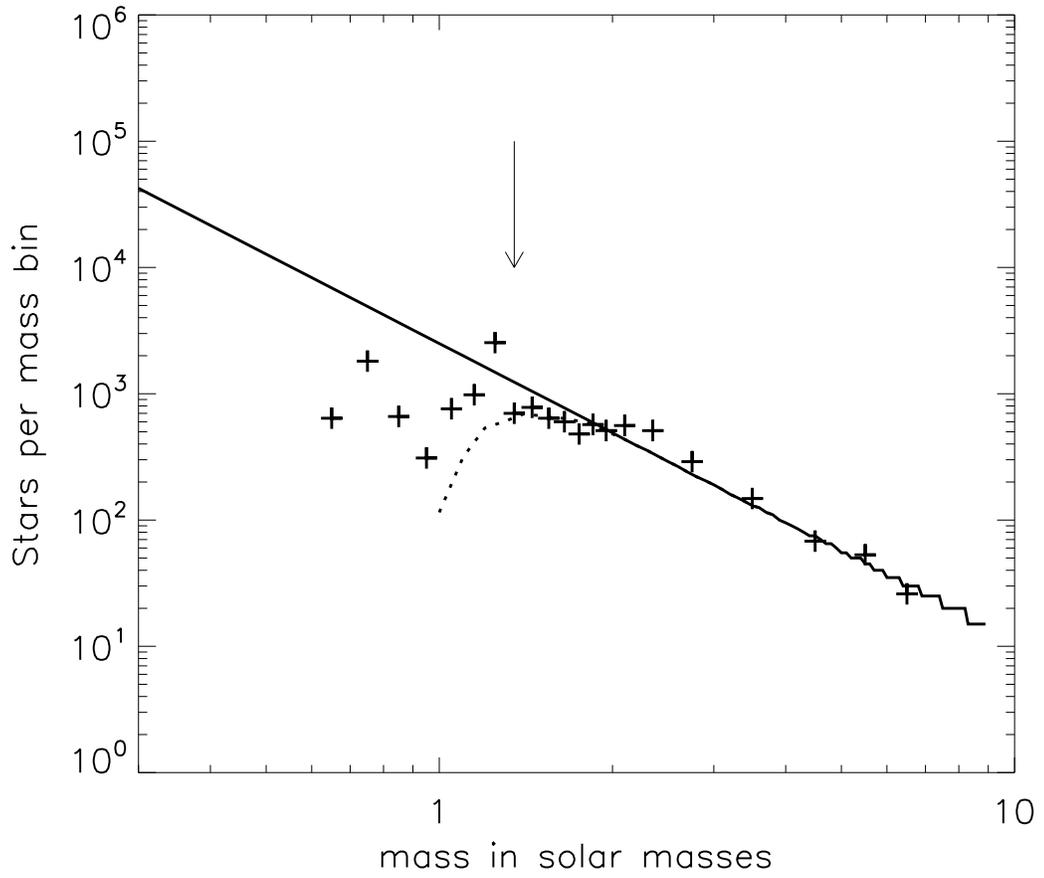}
\caption{An illustration of the effect on the derived IMF of individual 
random differential extinction 
within 30 Dor. 
The solid line is an artificial Salpeter mass function and the dotted line is 
the retrieved mass function by adopting a magnitude--limited sample. 
The magnitude limit was assumed to be the same as used by \citet{sirianni} 
(\filter{V}=24.7 mag). 
Crosses show the mass function derived by \citet{sirianni}. 
The stair-case shape of the input IMF at the high--mass end is due to  
finite, yet constant size of the mass bins. 
The 50\% completeness limit at 1.35 M$_\odot$ is indicated by the arrow. }
\label{checksirianni}
\end{center}
\end{figure}

Figure~\ref{checksirianni} shows the input Salpeter IMF (solid line), the 
derived IMF (dashed line) together with the measurements by \citet{sirianni}. 
The model mimics a flattening in the observed IMF similar to that deduced 
by \citet{sirianni}. 
The ratio of the number of stars below and above 2 M$_\odot$ respectively  
has been calculated both for the model cluster and the data 
from \citet{sirianni}. 
For the model cluster it is found to be 0.87, which is in reasonable 
agreement with the ratio of  0.76 derived from  the observations.


\subsection{Cumulative mass functions in the outer parts of R136}
Another  explanation  for the difference  between the results obtained here 
and the results by \citet{sirianni} can be mass segregation. 
We have searched for evidence for mass segregation in the two outer annuli 
in our survey. 
We  used the luminosity functions instead of the mass functions to avoid 
additional uncertainties due to the mass--luminosity relation. 
The results obtained for the mass functions are very similar to those from 
the luminosity functions. 

\begin{figure}
\begin{center}
\plotone{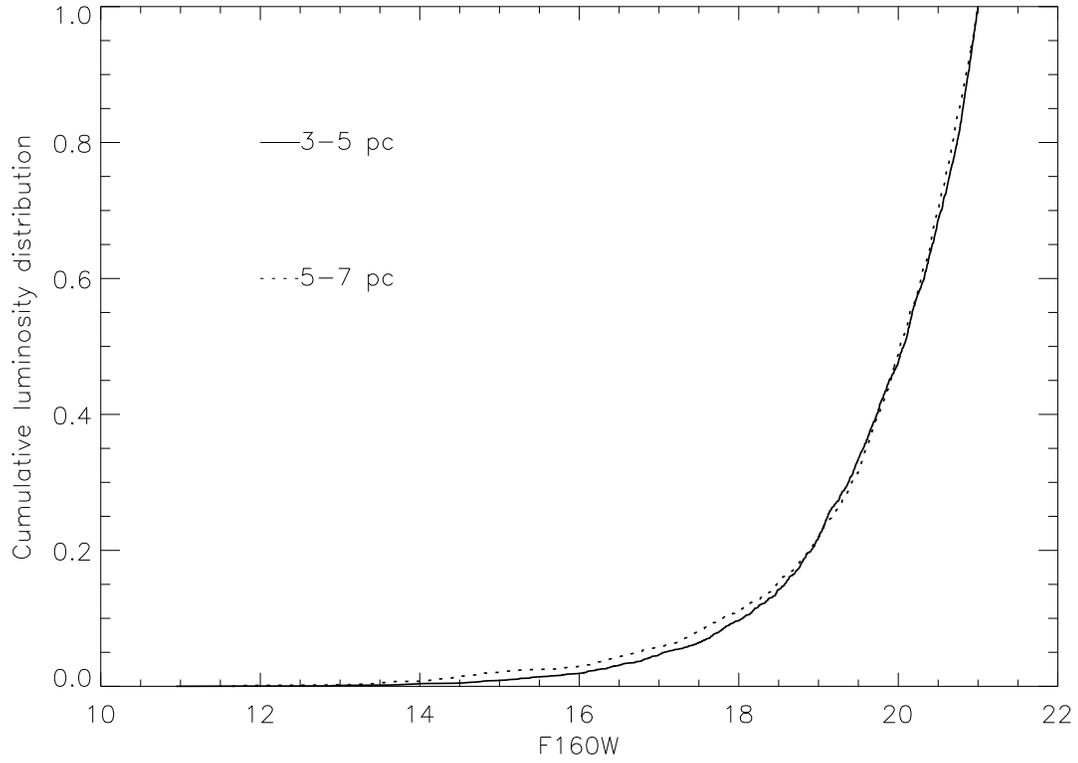}
\caption{A comparison oftwo cumulative luminosity distributions outside 
3 pc and down to the 50\%\ completeness level of \fw = 21 mag for the 3--5 
pc annulus. 
A K--S test of the two distributions gives a maximum distance between the 
two distributions of 0.039 and the probability the two distributions to be 
drawn from the same parent distribution is 10\%. 
The implication is that there is no evidence for luminosity segregation in 
the two annuli. }
\label{cumu_LF}
\end{center}
\end{figure}

The cumulative luminosity distributions are shown in Fig.~\ref{cumu_LF} 
for the outer two radial bins. These are the only bins where the 50\% 
completeness limit is below 2 M$_\odot$. The two cumulative distributions 
are very similar. 
We have performed a Kolmogorov--Smirnov test to quantify the similarity of the cumulative luminosity distributions. 
The maximum difference between the two distributions is 0.039 and the probability for the two distributions to be drawn from the same parent distribution is 10\%.
Thus, there is no strong evidence (less than 2$\sigma$) for mass segregation in the outer parts  of the cluster.

The fact that there is little evidence for mass segregation outside 3 pc does 
not exclude the possibility that the cluster is mass segregated out to a 
radius of several pc. 
Both \citet{malumuth} and \citet{brandl96} found evidence for mass segregation 
of the massive stars in the center of the cluster. 
\citet{brandl96} showed the half--mass relaxation time to be 7.8$\cdot10^7$ 
yr, much longer than the cluster age. 
They also point out that the massive stars will experience mass segregation 
on a much shorter time scale than the lower mass stars; the time scale 
depends inversely on the stellar mass. 
It is thus not surprising, from a dynamical point of view, that there is 
no evidence for mass segregation  outside the half--mass radius of 
1.7 pc \citep{hunter95}. 

On the other hand, this does not rule out the possibility that the cluster 
might be mass segregated at birth closer to the cluster center. 
Evidence for  mass segregation has been found in e.g. the Orion Nebula 
Cluster (ONC) \citep{hillenbrandhartmann,bonnell}. 
\citet{hillenbrandhartmann} showed evidence for mass segregation down to 
stellar masses of 1--2 M$_\odot$. 
Due to the youth of the ONC, they concluded the mass segregation had to be at 
least partly primordial. 
It is thus possible that R136 is also affected by primordial mass segregation 
close to the cluster center and that  mass segregation is the reason for the 
difference between the NICMOS and WFPC2 IMFs.

\subsection{Cluster mass}

We can obtain a rough estimate of the cluster mass from the near--infrared 
observations. 
The main limitation in our mass estimate is the amount of confusion due to 
crowding in the cluster centre: our data mainly samples the IMF down to and 
 below 1.4  M$_\odot$ outside 3 pc. 
Nevertheless, we can utilise the mass estimates within 2 pc from 
\citet{hunter95} to complement our mass estimate down to 2.1 M$_\odot$.  
 Inside 2 pc and into 0.15 pc, the results are extrapolated from the local 
 completeness limit mass down  to 2.1 M$_\odot$ assuming an 
underlying Salpeter IMF\@. 
No stars have been detected less massive than 20 M$_\odot$ within the 
central 0.15 pc radius due to crowding. 
The mass in the very center has been estimated from the surface density 
profile down to 2.8 M$_\odot$ in \citet{hunter95} to be 
$4\cdot10^4$ M$_\odot$pc$^{-2}$, resulting in a mass of 3700 M$_\odot$ 
down to a lower mass limit of 2.1 M$_\odot$.  
We find the cluster  total mass down to 2.1 M$_\odot$ to be 
$5\cdot10^4$ M$_\odot$. 
The directly determined mass down to 2.8 M$_\odot$ within 4.7 pc is found to 
be   $2.0\cdot10^4$ M$_\odot$, almost the same as found by \citet{hunter95}. 
If the IMF follows a Salpeter slope down to 0.5 M$_\odot$ as observed in the 
Galactic field and nearby lower--mass clusters \citep{kro02}, the total mass 
in the central region would be  roughly double the  amount given above, and 
the total cluster mass would be close to $\sim10^5$M$_\odot$. 

The velocity dispersion, and hence the dynamical mass, of the whole NGC 
2070 region, including R 136 has been determined by \citet{bosch}. 
The dynamical mass was determined to be 4.5$\cdot10^5$M$_\odot$, almost 5 
times higher than expected for R 136 alone, but consistent with the 
photometric mass for the same area \citep{selman2}. 
If we take into account that the half mass radius of R 136 is 1.7 pc 
\citep{hunter95}, compared to 14 pc for the whole NGC 2070 region and 
assuming the velocity dispersion is the same in the inner parts of the  
cluster, we would expect a dynamical mass of 
4.5$\cdot10^5\cdot 1.7/14$M$_\odot$=5.5$\cdot10^4$M$_\odot$ which is lower 
than the mass expected if the IMF is consistent with a Galactic IMF down to 
0.5 M$_\odot$. 
Thus, at face value, the velocity dispersion would be low enough that the 
cluster can stay bound. 
However, a measurement of the velocity dispersion for the inner regions is 
necessary to directly compare the photometric mass with the dynamical mass. 


\subsection{ The surface brightness profile} 
\begin{figure}[t]
\begin{center}
\plotone{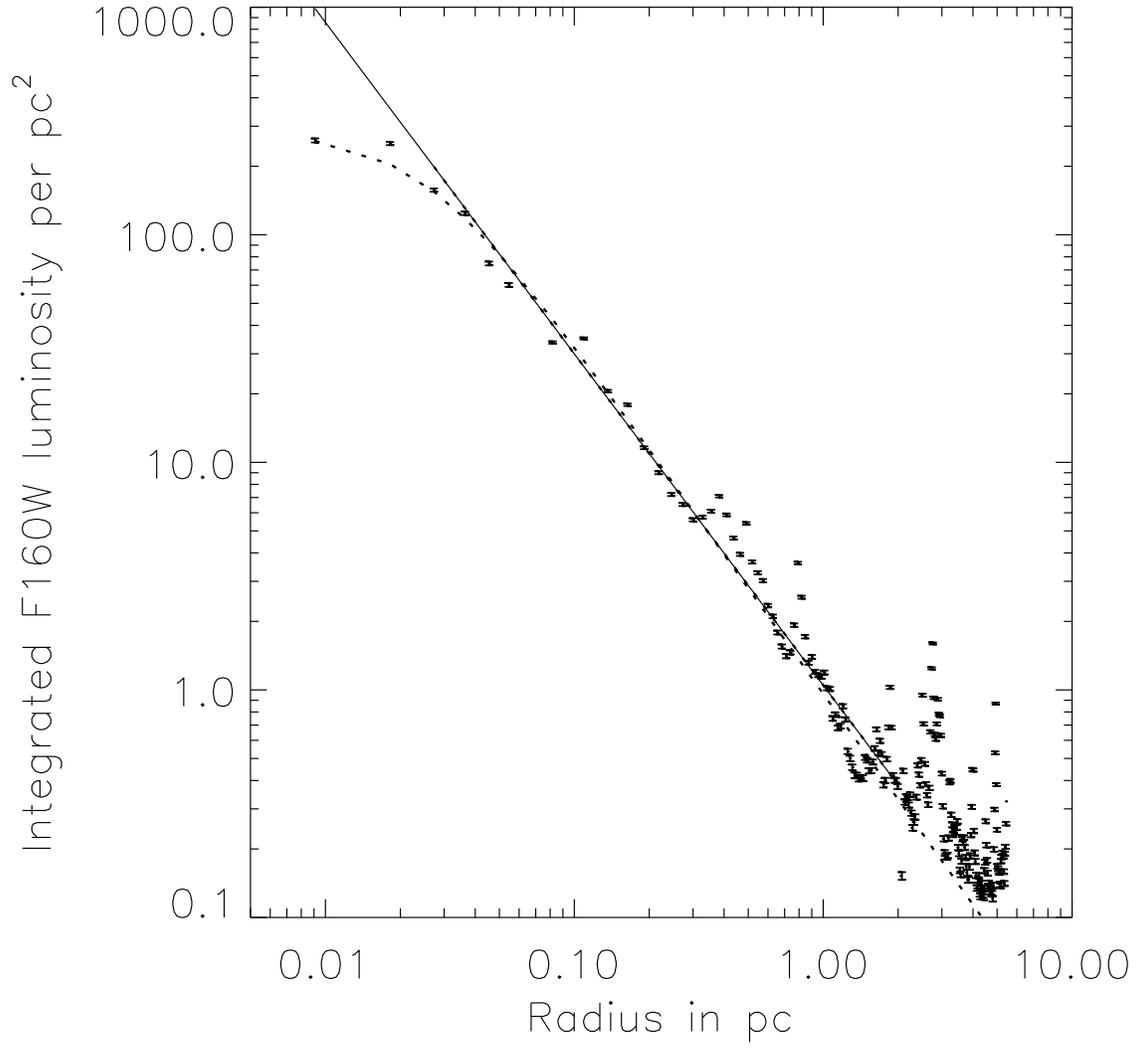}
\caption{The surface brightness profile of the R136 cluster within a 7 pc 
radius of  the cluster center. 
The solid line is a pure power--law fit from 0.02 pc to 2 pc with a derived slope
of $-1.48$. 
The dashed line is an Elson, Fall \& Freeman type profile with a core radius of 0.025pc and a power--law slope of $-1.54$. The fit was done from 0.009pc--2pc. 
Beyond 2 pc, the presence of the individual bright stars labeled in 
Fig.~\ref{30dormos} introduces the jitter seen in the surface 
brightness profile. }
\label{lightprof}
\end{center}
\end{figure}

We can  directly derive  the surface brightness  profile of the region around 
R136 in the 30 Dor cluster since the data does not suffer from saturated stars. 
Although bright stars will saturate through the one hour exposure, the 
non-destructive readout mode ensures that only the first reads are used to  
derive the magnitude of the brightest stars. 
The surface brightness profile is  shown in  Fig.~\ref{lightprof}. 

Between $\sim$0.2 and  2 pc, the light profile is well fit by a power--law, whereas inside 0.2 pc the light profile appears to be flattening. 
We have therefore fitted the light profile with a power--law modified by a core radius, similar to the approach in \citet{EFF}. 
Constraining the fit to inside 2 pc, we find a slope of $-1.54\pm0.02$, slightly more shallow than  $-1.72\pm0.06$ derived outside 0.1pc by \citet{campbell} using  \filter{F336W} Planetary Camera onboard HST observations.  

The core radius is found to be $0.025\pm0.004$pc, which is less than the resolution of the observations and is thus likely and upper limit. 
Previous HST optical studies determined a small core radius, $r_c\le 0.02$ pc 
\citep{hunter95}, consistent with our findings here. 
However, since the derived core radius is smaller than the resolution of the observations, it's evidence is weak. 
One or two bright stars off center by only a small amount could mimic a cluster core.

\subsection{Comparison with other massive clusters and the implications of low--mass stars in R136}



How does the low--mass end of the IMF in 30 Dor compare with that determined 
for other massive and dense stellar clusters? 
A top--heavy IMF in  massive dense clusters has  been suggested on theoretical 
grounds  \citep[e.g.][]{silk}. 
The most  convincing example of a young cluster with a  present-day mass 
function departing significantly from  a  Salpeter IMF above 1 M$_\odot$ is the 
Arches cluster \citep{stolte,figer}. 
\citet{stolte} found an average slope of $\Gamma=-0.9\pm0.15$ for the central 
parsec of the Arches cluster, flatter than a Salpeter slope of -1.35. 
Deeper observations found that the present day mass function in Arches 
to be well approximated by a power--law with a slope of $\Gamma=-0.91\pm0.08$ 
down to 1.3 M$_\odot$ \citep{kim}. 
However, recent work taking differential extinction into account suggest the slope of the power--law is only slightly more shallow than a Salpeter slope, $\Gamma=-1.1\pm0.2$ \citep{espinoza}. 
\citet{portegies} note that even if the observed IMF is slightly flatter thann a Salpeter IMF, this can be explaned by mass segregation. 
The mass segregation would be accelerated in the cluster due 
to the strong gravitational field from the Galactic Center. 
By adopting realistic parameters for a model cluster and an appropriate 
distance from the Galactic center, they found that an input Salpeter slope 
IMF would be transformed  to the observed present day mass function via 
strong  dynamical evolution. 
\citet{stolte2} showed that the IMF of the cluster powering the NGC 3603 HII 
region was well fitted by a power--law but with a  slope flatter than 
Salpeter, $\Gamma=-0.91\pm0.15$. 
They further showed evidence for  mass segregation for the more massive 
stars, M$> 4$ M$_\odot$. 
The data  indicated a slight flattening of the low--mass content 
(M$<$ 3 M$_\odot$). 
NGC 3603 is younger than the Arches cluster and not affected by a strong 
tidal gravitational field. Thus it is expected to be less 
influenced by dynamical 
mass segregation. 

The even more massive starburst clusters appear to be the primary sites 
(unit cells) 
of star formation in starburst galaxies, including 
 interacting/colliding galaxies such as The Antennae or The Cartwheel. 
If starburst clusters are the basic building blocks of certain star--forming 
galaxies, their stellar content (IMF) will affect much of the observed 
chemical and photometric evolution of galaxies, both at the 
present epoch and perhaps even more so in the high-redshift 
past \citep{charlot}. 
Several observational claims have been made that the IMF in unresolved 
starburst clusters is top--heavy \citep{rieke}, although observations of the 
Antennae gave a mixed result \citep{mengel}. 
However, it  has  been suggested that the high mass--to--light ratios found 
in some young starburst clusters are artificially high related to their
not being in virial equilibrium due to gas expulsion 
from the clusters \citep{goodwinbastian}. 
During the first 50 Myr of the cluster, the velocity dispersion and hence the 
cluster mass might be overestimated if the cluster is assumed to be virialized. 
\citet{goodwinbastian} suggest that the  top--heavy IMFs inferred in young 
unresolved extragalactic star clusters  
might be spurious due to their non-virialized dynamical state.

With the present dataset it is clear that the IMF in the outer parts of R136 
continues as a power--law down to 1 M$_\odot$, similar to what is found in 
other star clusters and the slope is similar to what is found in the field. 
Whether this is true for the cluster as a whole depends on the cause for the 
flattening observed closer to the cluster center. 
It would be interesting to know the IMF if the observations could 
 be  extended closer to the characteristic mass where the 
Galactic field star IMF flattens \citep[0.5 M$_\odot$][]{kro02}, a mass that 
can be reached in massive young clusters ($\le$ 4 Myr) in the LMC 
with  AO systems.

It has long been suggested R136 might be a proto--globular cluster 
\citep{meylan,larson93}. 
The question has been whether R136 would remain bound over a Hubble time. 
One consequence of a top--heavy IMF is that the cluster would dissolve 
soon after gas expulsion and mass loss due to evolution of the 
high--mass stars.   
However, the detection of  stars in R136 less massive than 1 M$_\odot$ 
gives the first {\it direct} evidence that low--mass stars are formed in a 
starburst  cluster. 
The fact that the IMF in the outer parts of R136 appears to be a Salpeter IMF 
down to at least 1 M$_\odot$ gives support to the notion the cluster might be 
a proto--globular cluster, albeit a light one. 
Early  gas expulsion and subsequent mass loss through stellar evolution  will 
disrupt star clusters deficient in low--mass stars during the first 5 Gyr of 
the clusters life \citep{chernoff,goodwin_old}
However, a determination of the velocity dispersion in the inner parts of the 
cluster is necessary to determine its final fate.  
Thus, the presence of low--mass stars is a necessary, but not sufficient 
condition for the  possibility of the cluster to evolve into a globular cluster. 
The median mass of Galactic globular clusters is 8.1$\cdot10^4$ M$_\odot$ 
\citep{mandushev}, comparable to the mass of R136.  
Even if R 136 will remain bound it will lose some mass and might end up 
as a low--mass globular cluster.

\section{Conclusions}
We have  analyzed HST/NICMOS  \filter{F160W} band data covering the central 
14pc$\times$14.25pc  around R136 in  the NGC 2070  cluster in the LMC. 
We have reached the following conclusions: 
\begin{itemize}
\item{From the color--magnitude diagram obtained by combining our photometry 
with previously published HST/WFPC 2 \filter{F555W} data  we constrain the age 
of the lower--mass stellar content in the cluster to be 2--4 Myr, consistent 
with previous estimates. 
We derive individual masses for the objects detected adopting a 3 Myr  
isochrone.}
\item{We have detected stars in  the cluster down  to  0.5 M$_\odot$  at   
$r > 5$ pc, assuming an age of 3 Myr.}
\item{The derived IMF is consistent with a Salpeter slope IMF with no 
evidence for a flattening at low masses down to the 50\%\ completeness limit 
corresponding to a mass  of 1.1 M$_\odot$  outside a radius of 5 pc for a 
3 Myr population and 1.4 M$_\odot$ if the oldest stars are 4 Myr. }
\item{The result is in disagreement with the flattening of the IMF below 2 
M$_\odot$ observed by \citet{sirianni} using optical data covering a region 
closer to the cluster center. 
We suggest two possible reasons for the discrepancy: differential extinction 
and mass segregation.}
\item{We find no evidence for mass segregation outside 3 pc, but with the 
current data, we cannot rule out that closer to the center the low--mass 
stars are segregated.}  
\item{From the radial surface brightness  profile we have derived a  core 
radius for the cluster of 0.025 pc (0\farcs1),  consistent with previous 
estimates by \citet{hunter95}.}
\item{The mass of the cluster within 7 pc between 25 M$_\odot$ and down to 
2.1 M$_\odot$ is estimated to be 5$\cdot10^4$ M$_\odot$. If the IMF continues 
with  a Salpeter slope down to 0.5 M$_\odot$ the total  mass estimate will 
double.}
\item{The total mass of the cluster combined with the  large number of 
low--mass stars suggests that  the 30 Dor cluster may survive to become   a 
proto--globular cluster depending on the cluster velocity dispersion.}
\end{itemize}

\acknowledgements
We thank Richard Larson for discussions in the early phases of the project, 
Eddie Bergeron for assistance with the drizzle software,  and Matthew 
Kenworthy for commenting  on an early version of the manuscript. 
M. A. and H. Z. acknowledges support from the DLR grant 50OR9912: 
``Data analysis of NICMOS/HST images of the 30 Dor cluster'' and partial 
funding through the DLR grant 50OR0401. 
M.A thanks the Astrophysikalisches Institut Potsdam for providing a 
stimulating and supportive environment for carrying out this Thesis work. 
Additional support was funded through the European Commission  Fifth Framework 
Programme Research Training Network ``The Formation and Evolution of Young 
Stellar Clusters'' (HPRN-CT-2000-00155). 
The Astronomische Gesellschaft  is acknowledged for providing funding for 
travel. 
Support for this work was provided by NASA through grant number 
GO-07370.01-96A from the Space Telescope Science Institute, which is 
operated by the Association of Universities for Research in Astronomy, 
Inc., under NASA contract NAS 5-26555.

Facilities: \facility{
This paper is based on observations made with the NASA/ESA 
{ \it Hubble Space Telescope}, operated by the Space Telescope Science 
Institute, which is operated by the Association of Universities for 
Research in Astronomy, Inc., under NASA contract NAS5-26555.}


\clearpage

\clearpage

\clearpage
\clearpage

\clearpage
\clearpage
\clearpage
\clearpage

\clearpage
\clearpage
\clearpage
\clearpage
\clearpage
\clearpage

\clearpage
\clearpage

\clearpage
\clearpage

\clearpage
\begin{table}
     \begin{displaymath} 
         \begin{tabular}{ccc}
            \hline
            \noalign{\smallskip}
$\mathrm{Annulus(pc)}$ & $\mathrm{slope} (\alpha)$ & $\mathrm{Max} (F160W)$ \\
$0.6-1.0$ & $0.38\pm0.07$ & $18.0$ \\
$1.0-2.0$ & $0.32\pm0.03$ & $19.5$\\
$2.0-3.5$ & $0.31\pm0.02$ & $20.5$\\
$3.5-5.0$ & $0.32\pm0.01$ & $21.0$\\
$5.0-7.0$ & $0.31\pm0.01$ & $21.5$ \\
\hline
         \end{tabular}
     \end{displaymath} 
\caption{The derived slopes ($\frac{dN}{dF160W}\propto10^{\alpha F160W}$) of 
the \fw\ band luminosity functions for R136 together with their uncertainties. 
The inner and outer radii are given for each annulus. 
The maximum \fw\ band magnitude (50\%\ completeness limit)  used to derive 
the fit is shown as well. }
\label{HLF_slopes}
   \end{table}

\begin{table}
     \begin{displaymath} 
         \begin{tabular}{cccc}
            \hline
            \noalign{\smallskip}

Xcen & 	Ycen & \filter{F160W} & $\filter{F160W}_{error}$\\
 104.546 & 280.899 & 19.09 & 0.04\\
156.263 & 148.125 & 20.28 & 0.05\\
121.580 & 148.709 & 20.15 & 0.03\\
315.194 & 150.819 & 20.29 & 0.03\\
481.127 & 152.319 & 20.27 & 0.04\\
69.890 & 154.680 & 17.73 & 0.05\\
280.115 & 155.118 & 19.76 & 0.01\\
91.993 & 157.119 & 18.87 & 0.04\\
443.582 & 157.392 & 17.34 & 0.02\\

\hline
         \end{tabular}
     \end{displaymath} 
\caption{The \filter{F160W} band detected sources in R 136. 
The full table is available online. 
}
\label{sources}
   \end{table}

\begin{table}
     \begin{displaymath} 
         \begin{tabular}{ccccccc}
            \hline
            \noalign{\smallskip}
$\mathrm{Age (Myr)}$ &$\mathrm{Annulus (pc)}$ & $\mathrm{Mass\  range (M}_\odot\mathrm{)}$ & $\mathrm{Slope}(\Gamma)$ & $\mathrm{Mass\  range (M}_\odot\mathrm{)}$ & $\mathrm{Slope}(\Gamma)$ \\
\hline
\noalign{\smallskip}
$3$ & $0.6-1.0$ & $8.9-20$ & $-1.7\pm0.3$ & -- & -- \\
$3$ & $1.0-2.0$ & $8.9-20$ & $-1.5\pm0.1$ & -- & -- \\
$3$ & $2.0-3.0$ & $8.9-20$ & $-1.4\pm0.2$ & -- & -- \\
$3$ & $3.0-5.0$ & $1.4-20$ & $-1.2\pm0.1$ & --  &  --  \\
$3$ & $5.0-7.0$ & $0.8-20$ & $-0.9\pm0.1$ & 1.4--1.7 & $-0.9\pm0.2$\\ 
$2$ & $3.0-5.0$ & $1.1-20$ & $-1.0\pm0.1$ & 1.1--1.7 & $-1.3\pm0.3$\\ 
$2$ & $5.0-7.0$ & $0.7-20$ & $-1.0\pm0.1$ & 0.7--1.7 & $-0.8\pm0.2$\\ 
$4$ & $3.0-5.0$ & $1.4-20$ & $-1.2\pm0.1$ & -- & -- \\ 
$4$ & $5.0-7.0$ & $1.1-20$ & $-1.2\pm0.1$ & 1.1--1.6 & $-1.3\pm0.4$\\ 
$2-4$ & $3.0-5.0$ & $1.4-20$ & $-1.3\pm0.1$ & -- & --\\ 
$2-4$ & $5.0-7.0$ & $1.1-20$ & $-1.2\pm0.1$ & 1.1--1.6 & $-1.1\pm0.4$\\ 

            \noalign{\smallskip}
            \hline
            \noalign{\smallskip}
            \hline
         \end{tabular}
     \end{displaymath} 
\caption{The derived slopes ($\Gamma$, $dN/d\log M\propto M^{\Gamma}$) of the 
mass functions together with the derived uncertainties of the  slopes. 
The slope of a Salpeter IMF in these units is $-1.35$. 
All the fits were performed over the mass range indicated in the table.}
\label{slopes}
   \end{table}

\begin{table}
     \begin{displaymath} 
         \begin{tabular}{cccccccccccccccc}
            \hline
Center of mass bin (M$_\odot$) &  0.7 & 0.9 & 1.1 & 1.4 & 2.0 & 2.8 & 3.5 & 5.0 & 7.1 & 8.9 &11.2 &14.1 &17.8 &22.4 &28.2\\
$\mathrm{Annulus(pc)}$ \\
\hline

0.6-1.0 &  0 &  0 &  0 &  0 &  2 & 11 & 13 & 21 & 42 & 51 & 50 & 25 & 18 & 11 &  8\\
1.0-2.0 &  9 & 27 & 44 & 56 & 92 &155 & 85 & 56 &111 & 91 & 77 & 48 & 35 & 30 & 10\\
2.0-3.0&  40 & 74 &126 &139 &163 &157 &100 & 53 &105 & 80 & 55 & 57 & 30 & 17 &  9\\
3.0-5.0 & 283 &328 &416 &380 &329 &268 &137 & 86 &168 &111 & 73 & 41 & 40 & 29 &  9\\
5.0-7.0 &310 &351 &355 &297 &278 &202 &102 & 56 &108 & 74 & 47 & 43 & 24 & 24 &  6\\
\hline
         \end{tabular}
     \end{displaymath} 
\caption{The number of stars per mass bin used to derive the IMF for the 
3 Myr isochrone. 
The numbers have not been corrected for incompleteness. }
\label{numbers}
   \end{table}

\end{document}